\documentclass[prd,showpacs]{revtex4}
\usepackage{epsf}
\usepackage{latexsym}

\usepackage{amsmath}
\usepackage{amssymb}
\usepackage{ifthen}

\newcounter{fig}

\newcommand{\vphi}{\varphi}
\newcommand{\DS}{\displaystyle}

\newcommand{\Psiel}{\tilde{\Psi}}
\newcommand{\psiel}{\tilde{\psi}}

\newcommand{\hD}{\hat{D}}

\begin{document}

\title{
Stationary Dyonic Regular and Black Hole Solutions}
\vspace{1.5truecm}
\author{Burkhard Kleihaus}
\affiliation{ZARM, Universit\"at Bremen, Am Fallturm, 
D--28359 Bremen, Germany}
\author{Jutta Kunz, Francisco Navarro-L\'erida, Ulrike Neemann}
\affiliation{Institut f\"ur Physik, Universit\"at Oldenburg, Postfach 2503,
D--26111 Oldenburg, Germany}

\date{\today}

\begin{abstract}
We consider globally regular and black hole solutions in SU(2) 
Einstein-Yang-Mills-Higgs theory, coupled to a dilaton field.
The basic solutions represent magnetic monopoles,
monopole-antimonopole systems or black holes with monopole or dipole hair.
When the globally regular solutions carry additionally electric charge,
an angular momentum density results, except in the simplest
spherically symmetric case.
We evaluate the global charges of the solutions and their effective action,
and analyze their dependence on the gravitational coupling strength.
We show, that in the presence of a dilaton field,
the black hole solutions satisfy a generalized Smarr type mass formula.
\end{abstract}


\maketitle

\section{Introduction}

In Einstein-Maxwell (EM) theory the Kerr-Newman (KN) solutions
represent stationary asymptotically flat black holes,
characterized uniquely by their global charges:
their mass $M$, their angular momentum $J$,
their electric charge $Q$, and their magnetic charge $P$
\cite{nohair1,nohair2}.
Following Wheeler this uniqueness theorem of EM theory is often
expressed as ``EM black holes have no hair''.

In many unified theories, including string theory, dilatons appear.
When a dilaton is coupled to EM theory, 
this has profound consequences for the black hole solutions.
Not only do
charged static Einstein-Maxwell-dilaton (EMD) black hole solutions
exist for arbitrarily small horizon size \cite{emd},
but also the staticity theorem of EM theory \cite{wald-st} does not
generalize to EMD theory for arbitrary dilaton coupling constant $\gamma$:
at the Kaluza-Klein value $\gamma_{\rm KK}=\sqrt{3}$
stationary non-static black holes appear, whose horizon is
non-rotating \cite{Rasheed,KKN-c},
and beyond $\gamma_{\rm KK}$ even counterrotating black holes
arise, whose horizon angular velocity and global angular momentum
have opposite sign \cite{KKN-c}.

The EM uniqueness theorem, on the other hand, 
does not readily generalize to theories with
non-Abelian gauge fields coupled to gravity \cite{review}.
The hairy black hole solutions of SU(2) Einstein-Yang-Mills (EYM)
and Einstein-Yang-Mills-Higgs (EYMH) theory
possess non-trivial magnetic fields outside their regular event horizon
and are not uniquely characterized by their mass, their
angular momentum, their electric and magnetic charge
\cite{su2bh,gmono,kk,hkk,kkrot,kknrot,KKNmass}.
Futhermore, black hole solutions arise, which are static 
and not spherically symmetric, showing that Israel's theorem
\cite{nohair1,nohair2} does not
generalize to non-Abelian theories, either \cite{kk,hkk}.

The coupling to non-Abelian fields not only gives rise to
new types of black hole solutions, but also allows for globally
regular solutions, not present in EM theory either \cite{bm,gmono,review}.
These are stationary solutions with a spatially localized energy
density of the matter fields and a finite mass,
and are referred to as solitons when they are stable,
and sphalerons when they possess unstable modes.
The known globally regular solutions of EYM theory
represent sphalerons, 
whereas the globally regular magnetic monopoles of EYMH theory are solitons, 
whose topological charge is proportional to their magnetic charge \cite{tHooft}.
Besides magnetic monopoles EYMH theory contains a plethora of
further globally regular solutions, representing for instance 
monopole-antimonopole pairs, chains,
and vortex ring solutions \cite{KKS}.

It is an interesting question whether such globally
regular solutions can be endowed with rotation, like their
black hole counterparts can.
When EYM black holes start to rotate, the
time component of their gauge potential is excited, as expected.
Surprisingly, however, not only a magnetic moment is induced by the rotation
but also an electric charge \cite{pert,kkrot},
and this seems to preclude the existence of globally regular
rotating EYM sphalerons \cite{bizon,eugen}.

Globally regular EYMH solutions, on the other hand, can carry electric charge,
and the presence of a time component of the gauge potential
renders the solutions stationary.
Together the electric and magnetic fields then give rise to an
angular momentum density,
except in the spherically symmetric case
\cite{eugen,eugen2,ulrike}.
Still, globally regular EYMH solutions with a non-vanishing global
magnetic charge cannot rotate: their
angular momentum vanishes
\cite{eugen,eugen2,ulrike}.
But globally regular EYMH solutions with no global magnetic charge
do possess a finite angular momentum.
In fact, it is proportional to their electric charge \cite{eugen},
giving rise to a quantization condition for the angular momentum,
\begin{equation}
J = n Q (1-\varepsilon) 
\ , \ \ \ P = n \varepsilon 
\ , \ \ \ \varepsilon= \frac{1}{2}\left[ 1 - (-1)^m \right]
\ , \label{quant} \end{equation}
where $m$ and $n$ are two integers, characterizing the EYMH solutions
\cite{KKS}.

Here we derive a mass formula 
for the stationary globally regular EYMH solutions
in the presence of a dilaton.
Then we address the dependence of the global charges
and of the effective action of these solutions
on the gravitational coupling strength.
For a given type of solution, typically two branches of solutions arise, 
which bifurcate at a maximal value of the coupling, $\alpha_{\rm max}$.
For static solutions, the mass $M$ exhibits a ``spike''
at $\alpha_{\rm max}$ \cite{gmono,KKS}, since there the two branches 
must possess the same tangent w.r.t.~$\alpha$ \cite{peter}.
When stationary and rotating solutions are considered, in contrast,
the mass branches may exhibit a ``loop'',
when considered as a function of $\alpha$ \cite{ulrike}.
Here we show that for stationary and rotating solutions it is 
the effective action ${S}^{\rm eff}$ which
may only exhibit a ``spike'' in the vicinity of the
maximal value of the gravitational coupling constant.
We illustrate this qualitative different behaviour of the mass $M$
and the effective action ${S}^{\rm eff}$
for several sets of numerically constructed stationary
and rotating solutions.

Turning to black holes again, we recall, that
EM black holes satisfy the laws of black hole mechanics \cite{wald}
and the Smarr mass formula \cite{smarr}
\begin{equation}
M = 2 TS + 2 \Omega J + \psiel_{\rm el} Q + \psiel_{\rm mag} P
\ , \label{smarr1} \end{equation}
where $T$ represents the temperature of the black holes
and $S$ their entropy,
$\Omega$ denotes their horizon angular velocity,
and 
$\psiel_{\rm el}$ and $\psiel_{\rm mag}$ represent
their horizon electric and magnetic potential, respectively.

In the presence of a dilaton an equivalent mass formula 
for EMD black holes is \cite{KKNmass}
\begin{equation}
M = 2 TS + 2 \Omega J + \frac{D}{\gamma} + 2\psiel_{\rm el} Q
\ , \label{namass} \end{equation}
with dilaton charge $D$ and dilaton coupling constant $\gamma$.
Interestingly, this second form of the mass formula also holds for
the known non-Abelian black hole solutions of 
Einstein-Yang-Mills-dilaton (EYMD) theory \cite{KKNmass}.

Here we address stationary black holes of EYMH and 
Einstein-Yang-Mills-Higgs-dilaton (EYMHD) theory
\cite{Forgacs2}.
For these black holes the zeroth law of black hole
mechanics holds \cite{kknrot}, 
as well as a generalized first law \cite{heustrau}.
We derive a mass formula for EYMHD black holes,
based on the asymptotic expansion of the metric
and the matter fields.
The analytical mass formula represents a good criterion
for the quality of numerically constructed EYMHD black hole solutions,
also presented.

In section {II} we recall the SU(2) EYMHD action and the equations of
motion.
We discuss the stationary ansatz for the metric, the gauge potential,
the Higgs field and the dilaton field,
and we present the boundary conditions for globally regular
and black hole solutions.
In section {III} we address the physical properties of the solutions.
We present the asymptotic expansion at infinity and the expansion at
the horizon, needed to obtain the global charges and the
horizon properties of the solutions. 
We evaluate the mass, the angular momentum and the effective action
of the globally regular solutions in section {IV},
and discuss the dependence of these quantities on the
coupling constant $\alpha$. We illustrate these results
for a set of numerically constructed solutions.
We then derive the mass formula for the
stationary black hole solutions in section {V},
presenting also numerical results.
In section {VI} we present our conclusions.

\section{\bf Einstein-Yang-Mills-Higgs-dilaton solutions}

After recalling the SU(2) EYMHD action and the general
set of equations of motion,
we discuss the ansatz for the stationary non-Abelian 
globally regular and black hole
solutions.
The ansatz for the metric represents the
stationary axially symmetric Lewis-Papapetrou metric
in isotropic coordinates. The ansatz for the gauge potential
and the Higgs field includes two integers, $m$ and $n$,
related to the polar and azimuthal angles.
For monopole-antimonopole chains the integer $m$
counts the total number of poles on the symmetry axis,
while the integer $n$ gives the magnitude of the magnetic charge
of each pole.
As implied by the boundary conditions,
the stationary axially symmetric solutions
are asymptotically flat, 
and the black hole solutions possess a regular event horizon.

\subsection{\bf SU(2) EYMHD Action}

We consider the SU(2) Einstein-Yang-Mills-dilaton action
\begin{equation}
{\cal S}=\int \left ( \frac{R}{16\pi G} + {L}_M \right ) \sqrt{-g} d^4x \ ,
\ \label{action} \end{equation}
where $R$ is the scalar curvature, and the matter Lagrangian ${L}_M$ 
is given by
\begin{equation}
{L}_M=-\frac{1}{2}\partial_\mu \Psi \partial^\mu \Psi
 - \frac{1}{2} e^{2 \kappa \Psi } {\rm Tr} (F_{\mu\nu} F^{\mu\nu})
-\frac{1}{4} {\rm Tr} \left( D_\mu \Phi D^\mu \Phi \right)
-\frac{\lambda}{8} e^{-2 \kappa \Psi } {\rm Tr} 
 \left( \Phi^2 - v^2 \right)^2
\ , \label{lagm} \end{equation}
with dilaton field $\Psi$,
gauge field strength tensor $ F_{\mu \nu} = 
\partial_\mu A_\nu -\partial_\nu A_\mu + i e \left[A_\mu , A_\nu \right] $,
gauge field $A_\mu =  A_\mu^a \tau_a/2$,
Higgs field in the adjoint representation $\Phi = \tau^a \Phi^a$,
gauge covariant derivative 
$D_\mu = \nabla_\mu + i e \left[A_\mu , \cdot \ \right] $,
and Newton's constant $G$, dilaton coupling constant $\kappa$,
Yang-Mills coupling constant $e$,
Higgs self-coupling constant $\lambda$, and Higgs vacuum
expectation value $v$.

The nonzero vacuum expectation value of the Higgs field
breaks the non-Abelian SU(2) gauge symmetry to the Abelian U(1) symmetry.
The particle spectrum of the theory then consists of a massless photon,
two massive vector bosons of mass $M_W = e v$,
and a massive Higgs field $M_H = {\sqrt {2 \lambda}}\, v$.
In the limit $\lambda = 0$ the Higgs field also becomes massless.
The dilaton is massless as well.

Including a boundary term \cite{Gibbons:1976ue},
variation of the action with respect to the metric and the matter fields
leads, respectively, to the Einstein equations 
\begin{equation}
G_{\mu\nu}= R_{\mu\nu}-\frac{1}{2}g_{\mu\nu}R = 8\pi G T_{\mu\nu}
\  \label{ee} \end{equation}
with stress-energy tensor
\begin{eqnarray}
T_{\mu\nu} &=& g_{\mu\nu}{L}_M 
-2 \frac{\partial {L}_M}{\partial g^{\mu\nu}}
 \nonumber \\
  &=& \partial_\mu \Psi \partial_\nu \Psi
     -\frac{1}{2} g_{\mu\nu} \partial_\alpha \Psi \partial^\alpha \Psi
      + 2 e^{2 \kappa \Psi }{\rm Tr}
    ( F_{\mu\alpha} F_{\nu\beta} g^{\alpha\beta}
   -\frac{1}{4} g_{\mu\nu} F_{\alpha\beta} F^{\alpha\beta})
\nonumber\\
&+&
\frac{1}{2} {\rm Tr} \left( D_\mu \Phi D_\nu \Phi - \frac{1}{2} g_{\mu\nu}
 D_\alpha \Phi D^\alpha \Phi \right)
- \frac{\lambda}{8} g_{\mu\nu}  e^{-2 \kappa \Psi }{\rm Tr}
 \left( \Phi^2 - v^2 \right)^2
\ , \label{tmunu}
\end{eqnarray}
and the matter field equations,
\begin{eqnarray}
& & D_\mu(e^{2 \kappa \Psi } F^{\mu\nu}) =
  \frac{1}{4} i e \left[\, \Phi, D^\nu \Phi \, \right]  \ ,
\label{feqA} \end{eqnarray}
\begin{eqnarray}
& &\Box \Psi =
 \kappa e^{2 \kappa \Psi }
  {\rm Tr} \left( F_{\mu\nu} F^{\mu\nu} \right)  
 -\frac{\lambda}{4} \kappa e^{-2 \kappa \Psi } {\rm Tr}
 \left( \Phi^2 - v^2 \right)^2
\ ,
\label{feqD} \end{eqnarray}
where $\Box \Psi = \Psi_{;\sigma}^{\ \ ;\sigma}$, and
\begin{eqnarray}
& &D_\mu D^\mu \Phi = \lambda e^{-2 \kappa \Psi } {\rm Tr}
 \left( \Phi^2 - v^2 \right)  \Phi \ .
\label{feqH} \end{eqnarray}

\subsection{\bf Stationary Axially Symmetric Ansatz}

The system of partial differential equations, 
Eq.~(\ref{ee}), Eq.~(\ref{feqD}), Eq.~(\ref{feqA}),  and  Eq.~(\ref{feqH})
is highly non-linear and complicated. 
In order to generate solutions to these equations,
one profits from the use of symmetries,
simplifying the equations.

Here we consider solutions,
which are both stationary and axially symmetric.
We therefore impose on the spacetime the presence of
two commuting Killing vector fields, 
$\xi$ (asymptotically timelike) and $\eta$ (asymptotically spacelike).
Since the Killing vector fields commute,
we may adopt a system of adapted coordinates, 
say $\{t, r, \theta, \varphi\}$, such that 
\begin{equation}
\xi=\partial_t \ , \ \ \ \eta=\partial_{\varphi}
\ . \label{xieta} \end{equation}
In these coordinates the metric is independent of $t$ and $\varphi$. 
We also assume that the symmetry axis of the spacetime, 
the set of points where $\eta=0$, is regular,
and satisfies the elementary flatness condition
\begin{equation}
\frac{X,_\mu X^{,\mu}}{4X} = 1 \ , \ \ \
X=\eta^\mu \eta_\mu \
\ .  \label{regcond} \end{equation}

Apart from the symmetry requirement on the metric 
(${\cal L}_{\xi} g = {\cal L}_{\eta} g =0$, 
i.e., $g_{\mu \nu}=g_{\mu \nu}(r,\theta)$), 
we impose that the matter fields are also symmetric 
under the spacetime transformations generated by $\xi$ and $\eta$. 

This implies for the dilaton field
\begin{equation}
{\cal L}_{\xi} \Psi = {\cal L}_{\eta} \Psi =0
\ ,  \label{symd} \end{equation}
so $\Psi$ depends on $r$ and $\theta$ only.
Introducing two compensating su(2)-valued functions
$W_{\xi}$ and $W_{\eta}$, the 
concept of generalised symmetry \cite{Forgacs,eugen} requires
for the Higgs field
\begin{equation}
{\displaystyle {\cal L}_{\xi} \Phi = ie[ \Phi, W_{\xi}] \ , \ \ \
 {\cal L}_{\eta} \Phi = ie[ \Phi, W_{\eta} ] }
\ ,  \label{symh} \end{equation}
and for the gauge potential $A=A_\mu dx^\mu$, 
\begin{eqnarray}
\displaystyle ({\cal L}_{\xi}A)_{\mu} &=&  D_{\mu} W_{\xi}
\ , \nonumber \\
\displaystyle ({\cal L}_{\eta}A)_{\mu} &=&  D_{\mu} W_{\eta}
\ , \label{symA} \end{eqnarray}
where $W_{\xi}$ and $W_{\eta}$ 
satisfy
\begin{equation}
{\cal L}_{\xi}W_{\eta}-{\cal L}_{\eta}W_{\xi} 
                + i e \left[W_{\xi},W_{\eta}\right] = 0
\ .  \label{compat} \end{equation}
Performing a gauge transformation to set $W_{\xi}=0$, 
leaves $\Phi$, $A$ and $W_{\eta}$ independent of $t$.

By virtue of the Frobenius condition and the
circularity theorem,
the metric can then be written in the Lewis-Papapetrou 
form, which in isotropic coordinates reads
\begin{equation}
ds^2 = -fdt^2+\frac{m}{f}\left[dr^2+r^2 d\theta^2\right] 
       +\sin^2\theta r^2 \frac{l}{f}
          \left[d\vphi-\frac{\omega}{r}dt\right]^2 \  
\ , \label{metric} \end{equation}
where $f$, $m$, $l$ and $\omega$ are functions of $r$ and $\theta$ only.

The $z$-axis represents the symmetry axis.
The regularity condition along the $z$-axis Eq.~(\ref{regcond}) requires
\begin{equation}
m|_{\theta=0,\pi}=l|_{\theta=0,\pi}
\ . \label{lm} \end{equation}

The event horizon of stationary black hole solutions
resides at a surface of constant radial coordinate, $r=r_{\rm H}$,
and is characterized by the condition $f(r_{\rm H},\theta)=0$ \cite{kkrot}.
The Killing vector field
\begin{equation}
\chi = \xi + \frac{\omega_{\rm H}}{r_{\rm H}} \eta
\ , \label{chi} \end{equation}
is orthogonal to and null on the horizon \cite{wald}.
The ergosphere, defined as the region in which $\xi_\mu \xi^\mu$ is positive,
is bounded by the event horizon and by the surface where
\begin{equation}
 -f +\sin^2\theta \frac{l}{f} \omega^2 = 0 \ .
 \label{ergo}
\end{equation}

For the gauge fields we employ a generalized ansatz \cite{kkrot,kknrot,ulrike}, 
which trivially fulfils both the symmetry constraints 
Eq.~(\ref{symA}) and Eq.~(\ref{compat}) 
and the circularity conditions,
\begin{equation}
A_\mu dx^\mu
  =   \left( B_1 \frac{\tau_r^{(n,m)}}{2e} + B_2 \frac{\tau_\theta^{(n,m)}}{2e}
 \right) dt 
+A_\varphi (d\varphi-\frac{\omega}{r} dt)
+\left(\frac{H_1}{r}dr +(1-H_2)d\theta \right)\frac{\tau_\varphi^{(n)}}{2e}
 , \label{a1} \end{equation}
\begin{equation}
A_\varphi=   -n\sin\theta\left(H_3 \frac{\tau_r^{(n,m)}}{2e}
            +(1-H_4) \frac{\tau_\theta^{(n,m)}}{2e}\right)  
 , \label{a2} \end{equation}
and the appropriate Ansatz for the Higgs field is then given by 
\cite{kknrot,ulrike}
\begin{equation}
\Phi =v \left( \Phi_1 \tau_r^{(n,m)} + \Phi_2 \tau_\theta^{(n,m)} \right)
\ , \label{a4} \end{equation}
where $n$ and $m$ are integers.
The symbols $\tau_r^{(n,m)}$, $\tau_\theta^{(n,m)}$ and $\tau_\vphi^{(n)}$
denote the dot products of the Cartesian vector of Pauli matrices,
$\vec \tau = ( \tau_x, \tau_y, \tau_z) $,
with the spatial unit vectors
\begin{eqnarray}
{\hat e}_r^{(n,m)} & = & \left(
\sin(m\theta) \cos(n\vphi), \sin(m\theta)\sin(n\vphi), \cos(m\theta)
\right)\ , \nonumber \\
{\hat e}_\theta^{(n,m)} & = & \left(
\cos(m\theta) \cos(n\vphi), \cos(m\theta)\sin(n\vphi), -\sin(m\theta)
\right)\ , \nonumber \\
{\hat e}_\vphi^{(n)} & = & \left( -\sin(n\vphi), \cos(n\vphi), 0 \right)\ ,
\label{unit_e}
\end{eqnarray}
respectively.
Like the dilaton field function $\Psi$,
the gauge field functions $B_i$ and $H_i$
and the Higgs field functions $\Phi_i$
depend only on the coordinates $r$ and $\theta$.

The ansatz is form-invariant under Abelian gauge transformations $U$
\cite{kk,kkreg}
\begin{equation}
 U= \exp \left({\frac{i}{2} \tau^{(n)}_\varphi \Gamma(r,\theta)} \right)
\ .\label{gauge} \end{equation}
With respect to this residual gauge degree of freedom 
we choose the gauge fixing condition 
$r\partial_r H_1-\partial_\theta H_2 =0$.
For the gauge field ansatz, Eqs.~(\ref{a1})-(\ref{a2}),
the compensating matrix $W_{\eta}$ is given by
\begin{equation}
W_{\eta}=  n\frac{\tau_z}{2 e}
\ . \label{Wcomp} \end{equation}

\subsection{\bf Dimensionless Quantities}

Let us now introduce the dimensionless quantities,
beginning with
the dimensionless coupling constants $\alpha$, $\beta$ and $\gamma$
\begin{equation}
v = \frac{\alpha}{\sqrt{4 \pi G}} \ , \ \ \
\lambda = e^2 \beta^2 \ , \ \ \
\kappa = \frac{\sqrt{4 \pi G}}{\alpha} \gamma
\ . \label{dimlesscon} \end{equation}
The dimensionless coordinate $x$ is given by
\begin{equation}
r=\frac{\sqrt{4\pi G}}{e \alpha}  x
\ , \label{dimless} \end{equation}
the dimensionless electric gauge field functions
${\bar B}_1$ and ${\bar B}_2$ are
\begin{equation}
{B}_1 = \frac{e \alpha}{\sqrt{4 \pi G}}  {\bar B}_1 \ , \ \ \
{B}_2 = \frac{e \alpha}{\sqrt{4 \pi G}}  {\bar B}_2 \ ,
\label{barb} \end{equation}
and the dimensionless dilaton function $\psi$ is
\begin{equation}
\Psi = \frac{\alpha}{\sqrt{4\pi G}} \psi
\ . \label{dimp} \end{equation}
Introducing these dimensionless quantities into the EOMs,
the resulting equations depend only on the parameters $\alpha$,
$\beta$, and $\gamma$.
Note, that in the limit $\gamma \rightarrow 0$
the dilaton decouples and the equations of EYMH theory are obtained. 

\subsection{\bf Boundary Conditions}

\noindent {\sl \bf Boundary conditions at infinity}

To obtain asymptotically flat solutions, we impose
on the metric functions
the boundary conditions at infinity 
\begin{equation}
f|_{x=\infty}= m|_{x=\infty}= l|_{x=\infty}=1 \ , \ \ \
\omega|_{x=\infty}= 0
\ . \label{bc1a} \end{equation}

For the dilaton function we choose
\begin{equation}
\psi|_{x=\infty}=0
\ , \label{bc1b} \end{equation}
since any finite value of the dilaton field at infinity
can always be transformed to zero via
$\psi \rightarrow \psi - \psi(\infty)$,
$x \rightarrow x e^{-\gamma \psi(\infty)} $.

The asymptotic values of the Higgs field functions $\Phi_i$ are
\begin{equation}
\Phi_1|_{x=\infty}= 1 \ , \ \ \
\Phi_2|_{x=\infty}= 0 
\ . \label{bc1c} \end{equation}

We further impose, that the two electric gauge field functions 
$\bar B_i$ satisfy
\begin{equation}
\bar B_1|_{x=\infty}= \nu \ , \ \ \ \bar B_2|_{x=\infty}= 0
 \ , \end{equation}
where the asymptotic value $\nu$ is restricted to
$0 \le |\nu| < 1$,
and that the magnetic gauge field functions $H_i$ satisfy
\begin{equation}
H_1|_{x=\infty}= 0 \ , \ \ \ H_2|_{x=\infty}= 1 - m 
\ , \end{equation}
\begin{equation}
H_3|_{x=\infty}=  \frac{\cos\theta - \cos(m\theta)}{\sin\theta}
\ \ \ m \ {\rm odd} \ , \ \ \
H_3|_{x=\infty}= \frac{1 - \cos(m\theta)}{\sin\theta}
\ \ \ m \ {\rm even} \ , \ \ \
\end{equation}
\begin{equation}
H_4|_{x=\infty}= 1- \frac{\sin(m\theta)}{\sin\theta} \ .
\end{equation}

\noindent {\sl \bf  Boundary conditions at the origin}

To obtain globally regular solutions, we must impose appropriate 
boundary conditions at the origin.
Regularity 
requires for the metric functions the boundary conditions
\begin{equation}
\partial_x f|_{x=0}=
\partial_x m|_{x=0}=
\partial_x l|_{x=0}= 0 \ , \ \ \
\omega|_{x=0}= 0
\ , \label{bc2a} \end{equation}
and for the dilaton function
\begin{equation}
\partial_x \psi|_{x=0}=0
\ , \label{bc2b} \end{equation}
the gauge field functions $H_i$ satisfy
\begin{equation}
H_1|_{x=0}= H_3|_{x=0}= 0\ , \ \ \ \
H_2|_{x=0}= H_4|_{x=0}= 1 \ ,
\end{equation}
while for even $m$ the gauge field functions $\bar B_i$ 
and the Higgs functions $\Phi_i$ satisfy 
\begin{equation}
\left. \left[ 
\sin(m\theta) \Phi_1 + \cos(m\theta) \Phi_2 
\right] \right|_{x=0} = 0 \ ,
\end{equation}
\begin{equation}
\left.\partial_x\left[\cos(m\theta) \Phi_1
                    - \sin(m\theta) \Phi_2 \right] \right|_{x=0} = 0 \ ,
\end{equation}
\begin{equation}
\left. \left[ 
\sin(m\theta) \bar B_1 + \cos(m\theta) \bar B_2 = 0 
\right] \right|_{x=0} = 0 \ ,
\end{equation}
\begin{equation}
\left.\partial_x\left[\cos(m\theta) \bar B_1
                    - \sin(m\theta) \bar B_2 \right] \right|_{x=0} = 0 \ ,
\end{equation}
whereas for odd $m$ they satisfy $\bar B_i|_{x=0}=\Phi_i|_{x=0}=0$.\\

\noindent {\sl \bf  Boundary conditions at the horizon}

The event horizon of stationary black hole solutions
resides at a surface of constant radial coordinate, $x=x_{\rm H}$,
and is characterized by the condition $f(x_{\rm H},\theta)=0$ \cite{kkrot}.

Regularity at the horizon then requires the following boundary conditions
for the metric functions
\begin{equation}
f|_{x=x_{\rm H}}=
m|_{x=x_{\rm H}}=
l|_{x=x_{\rm H}}=0
\ , \ \ \ \omega|_{x=x_{\rm H}}=\omega_{\rm H}= {\rm const.}
\ , \label{bh2a} \end{equation}
for the dilaton function 
\begin{equation}
\partial_x \psi|_{x=x_{\rm H}} =0
\ , \label{bh2b} \end{equation}
while the Higgs and the magnetic gauge field functions satisfy
\begin{equation}
\partial_x \Phi_1|_{x=x_{\rm H}}= \partial_x \Phi_2|_{x=x_{\rm H}}= 0
\ , \label{bh2c} \end{equation}
\begin{equation}
           H_1 |_{x=x_{\rm H}}= 0 \ , \ \ \
\partial_x H_2 |_{x=x_{\rm H}}= 
\partial_x H_3 |_{x=x_{\rm H}}= 
\partial_x H_4 |_{x=x_{\rm H}}= 0 \  
\ , \label{bh2d} \end{equation}
with the gauge condition $\partial_\theta H_1=0$ taken into account
\cite{kkrot}.
The boundary conditions for the electric gauge field functions
are obtained from the requirement that for non-Abelian
solutions the electrostatic potential 
is constant at the horizon \cite{kknrot}
\begin{equation}
\Psiel_{\rm el} \frac{\tau_z}{2}= - \chi^\mu A_\mu |_{r=r_{\rm H}}
\ . \label{esp0} \end{equation}
Defining the dimensionless electrostatic potential $\psiel_{\rm el}$,
\begin{equation}
{\psiel_{\rm el}} = \frac{\sqrt{4 \pi G}}{\alpha}  \Psiel_{\rm el} \ , 
\label{psi} \end{equation}
and the dimensionless horizon angular velocity $\Omega$,
\begin{equation}
 \Omega = \frac{\omega_{\rm H}}{x_{\rm H}}
\ , \label{Omega} \end{equation}
yields the boundary conditions 
\begin{equation}
\bar B_1 |_{x=x_{\rm H}}  =   n \Omega \cos m \theta  \ , \ \ \
\bar B_2 |_{x=x_{\rm H}}  =  -n \Omega \sin m \theta  \ .
\label{bh2e}
\end{equation}
\\

\noindent {\sl \bf Boundary conditions along the symmetry axis}

The boundary conditions along the $z$-axis
($\theta=0$ and $\theta=\pi$) are determined by the
symmetries.
For the positive $z$-axis they are given by
\begin{eqnarray}
& &\partial_\theta f|_{\theta=0} =
   \partial_\theta m|_{\theta=0} =
   \partial_\theta l|_{\theta=0} =
   \partial_\theta \omega|_{\theta=0} = 0 \ ,
\label{bc4a} \end{eqnarray}
\begin{eqnarray}
& &\partial_\theta \psi|_{\theta=0} = 0 \ ,
\label{bc4b} \end{eqnarray}
\begin{eqnarray}
 H_1|_{\theta=0}=H_3|_{\theta=0}=0 \ , \ \ \
   \partial_\theta H_2|_{\theta=0} =
   \partial_\theta H_4|_{\theta=0}  = 0 \ ,
\label{bc4c} \end{eqnarray}
\begin{eqnarray}
    \bar B_2|_{\theta=0}=0 \ , \ \ \ \partial_\theta \bar B_1|_{\theta=0}=0 \ ,
\label{bc4d} \end{eqnarray}
\begin{eqnarray}
    \Phi_2|_{\theta=0}=0 \ , \ \ \ \partial_\theta \Phi_1|_{\theta=0}=0 \ .
\label{bc4e} \end{eqnarray}
The analogous conditions hold on the negative $z$-axis.
We note, that the globally regular solutions are symmetric w.r.t.~the
$xy$-plane. For the black hole solutions, this symmetry is broken
via the boundary conditions of the time component of the gauge field.

In addition, regularity on the $z$-axis requires condition Eq.~(\ref{lm})
for the metric functions to be satisfied,
and regularity of the energy density on the $z$-axis requires
\begin{equation}
H_2|_{\theta=0}=H_4|_{\theta=0}
\ . \label{h2h4} \end{equation}

\section{\bf Properties of Regular and Black Hole Solutions}

We derive the properties of the stationary axially symmetric
solutions from the expansions of their metric and matter 
field functions at infinity, at the origin and at the horizon. 
The expansion at infinity yields
the global charges of the solutions,
the expansion at the horizon yields the horizon properties
of the black holes.

\subsection{\bf Asymptotic Expansion}

The asymptotic expansion depends on the integers $m$ and $n$.
Here we restrict to odd winding number $n$, since the
analysis for the even case seems to be `prohibitively complicated'.
We then obtain for $\beta=0$
\begin{eqnarray}
&&H_1 = -\frac{C_1 \sin\theta}{x} -\frac{C_3 \sin(2\theta)}{x^2} +
O\left(\frac{1}{x^3}\right) \ , \nonumber \\
&&H_2=(1-m) -\frac{C_1 \cos\theta}{x} - \frac{C_3 \cos(2\theta)}{x^2}+
O\left(\frac{1}{x^3}\right) \ ,\nonumber \\
&&H_3 = \frac{\cos(\varepsilon\theta) - \cos(m\theta)}{\sin\theta} + \frac{C_6
  \sin\theta}{x} + O\left(\frac{1}{x^2}\right) \ , \nonumber \\
&&H_4 = \left(1-\frac{\sin(m\theta)}{\sin\theta}\right) -\frac{C_1
  \cos(\varepsilon \theta)}{x} - \frac{C_3 \cos^2\theta + C_1 C_6 \sin^2\theta}{x^2} +
O\left(\frac{1}{x^3}\right) \ , \nonumber \\
&&\bar B_1 = \nu -\frac{Q}{x} + \frac{2Q (\mu-\gamma D) -\nu C_1^2 \sin^2\theta + 2
  C_7 \cos\theta}{2 x^2} + O\left(\frac{1}{x^3}\right) \ , \nonumber \\
&&\bar B_2 = \frac{\nu C_1 \sin\theta}{x} - \frac{(C_1 Q -\nu C_3
  \cos\theta)\sin\theta}{x^2} + O\left(\frac{1}{x^3}\right) \ , \nonumber \\
&&f = 1 -\frac{2 \mu}{x} + \frac{2 \mu^2 + \alpha^2 (Q^2+P^2) + C_4
  \cos\theta}{x^2} +  O\left(\frac{1}{x^3}\right) \ , \nonumber \\
&&m=1+\frac{C_5 \cos(2\theta) + [-\mu^2 +
  \alpha^2(Q^2+P^2-D^2-C_2^2)]\sin^2\theta}{x^2} + O\left(\frac{1}{x^3}\right)
\ , \nonumber \\
&&l = 1 + \frac{C_5}{x^2} + O\left(\frac{1}{x^3}\right) \ , \nonumber \\
&&\omega = \frac{2 \zeta}{x^2}  + O\left(\frac{1}{x^3}\right)
\ , \nonumber \\
&&\psi = -\frac{D}{x} - \frac{\gamma(Q^2-P^2) -2 C_8 \cos\theta}{2 x^2}  +
O\left(\frac{1}{x^3}\right) \ , \nonumber \\
&&\Phi_1 = 1 + \frac{C_2}{x} - \frac{C_1^2 \sin^2\theta -2 C_9 \cos\theta}{2
  x^2}  + O\left(\frac{1}{x^3}\right) \ , \nonumber \\
&&\Phi_2 = \frac{C_1 \sin\theta}{x} + \frac{(C_1 C_2 + C_3
  \cos\theta)\sin\theta}{x^2}  + O\left(\frac{1}{x^3}\right) \ ,
\end{eqnarray}
where 
\begin{equation}
\varepsilon = \frac{1}{2}[1-(-1)^m], \ \ \ P=n\varepsilon \ .
\end{equation}

For generic $\beta \neq 0$ solutions, the expansions remain valid with
$C_2=C_9=0$. 
At first sight the power law decay
of the Higgs field then appears surprising, since $\beta \neq 0$
renders the Higgs massive and should thus lead to an exponential decay.
However, this power law decay represents a gauge artifact and
can be removed by the gauge transformation
\begin{equation}
U= \exp(i \Gamma \tau_\vphi^n/2)
\ , \end{equation}
with
\begin{equation}
\Gamma =   -(1-m)\theta 
+\frac{C_1 \sin\theta}{x} +\frac{C_3 \sin(2\theta)}{2x^2}
\ . \end{equation}
Performing this gauge transformation 
leads to $\Phi = v \tau_r^{(n,1)} + O(1/r^3)$ 
(note that the integer $m$ has been transformed away).
We further obtain trivial gauge field functions
$H_1$ and $H_2$ (up to order $O(1/x^3)$).

\subsection{\bf Global Charges}

The expansion coefficients $M$, $J$, $Q$, $P$ and $D$
correspond to the global charges of the solutions.
The dimensionless mass $M$ and angular momentum $J$ of the solutions
are obtained from the asymptotic expansion of the metric
\begin{equation}
M = \frac{1}{2\alpha^2} \lim_{x \rightarrow \infty} x^2 \partial_x  f
 = \frac{\mu}{\alpha^2}
\ , \ \ \
J = \frac{1}{2\alpha^2} \lim_{x \rightarrow \infty} x^2 \omega
 = \frac{\zeta}{\alpha^2}
\ , \label{MJ} \end{equation}
These correspond to the expressions obtained from the 
respective Komar integrals,
as shown in sections IV and V
for the globally regular and black hole solutions, respectively.
Note that $M$ is the mass in units of $4 \pi v/e$,
whereas $\mu$ corresponds to the mass in units of Planck mass.
Likewise, the dimensionless electric charge $Q$ 
and the dimensionless magnetic charge $P$ are given by
\begin{equation}
Q = -\lim_{x \rightarrow \infty} x \left( \bar B_1 -\nu \right)
\ , \ \ \
P = \frac{n}{2}\left(1 - (-1)^m\right)
\ , \label{QP} \end{equation}
while the dimensionless dilaton charge $D$ is given by
\begin{equation}
D =  \lim_{x\rightarrow\infty}x^2\partial_x\psi
\ . \label{D} \end{equation}

\subsection{\bf Expansion at the Horizon}

Expanding the metric and matter field functions at the horizon
in powers of
\begin{equation}
\delta=\frac{x}{x_{\rm H}}-1
\end{equation}
yields to lowest order
\begin{eqnarray}
H_1 &=&\delta \left(1 -\frac{1}{2}\delta \right) H_{11}
+ O(\delta^3) \ , \nonumber\\
H_2 &=&H_{20}+O(\delta^2) 
\ , \nonumber\\
H_3 &=&H_{30} + O(\delta^2)
\ , \nonumber\\
H_4 &=&H_{40}+ O(\delta^2) \ , \nonumber \\
{\bar B}_1 &=&n\frac{\omega_{\rm H} }{x_{\rm H}}\cos{(m \theta)}
+ O(\delta^2) \ , \nonumber\\
{\bar B}_2 &=&-n\frac{\omega_{\rm H} }{x_{\rm H}}\sin{(m \theta)}
+ O(\delta^2) \ , \nonumber\\
f &=&\delta^2 f_2 (1 -\delta) + O(\delta^4) 
\ , \nonumber\\
m &=&\delta^2 m_2 (1 -3\delta) + O(\delta^4) 
\ , \nonumber\\
l &=&\delta^2 l_2 (1 -3\delta) + O(\delta^4) 
\ , \nonumber\\
\omega &=&\omega_{\rm H} (1 + \delta) + O(\delta^2) 
\ , \nonumber\\
{\psi} &=&\psi_0 + O(\delta^2) 
\ , \nonumber\\
\Phi_1 &=&\Phi_{10}  + O(\delta^2) \ , \nonumber \\
\Phi_2 &=&\Phi_{20}  + O(\delta^2) \ .
\label{H4_hor_red}
\end{eqnarray}
The expansion coefficients $f_2$, $m_2$, $l_2$, $\psi_0$, $H_{11}$,
$H_{20}$, $H_{30}$, $H_{40}$, $\Phi_{10}$, $\Phi_{20}$ are functions of the variable $\theta$.
Among these coefficients the following relations hold,
\begin{equation}
0=\frac{\partial_{\theta}m_2}{m_2}
-2 \frac{\partial_{\theta}f_2}{f_2}
\ , \label{relation_hor_1} \end{equation}
\begin{equation}
H_{11}=\partial_{\theta} H_{20}
\ . \label{relation_hor_2} \end{equation}

\subsection{\bf Horizon Properties}

With help of the above expansion
we obtain the horizon properties of the 
SU(2) EYMHD black hole solutions.
The first quantity of interest is the area of the black hole horizon. 
The dimensionless area $A_{\rm H}$ is given by
\begin{equation}
A_{\rm H} = 2 \pi \int_0^\pi  d\theta \sin \theta
\frac{\sqrt{l_2 m_2}}{f_2} x_{\rm H}^2
\ , \label{area} \end{equation}
and the dimensionless entropy $S$ by
\begin{equation}
S = \frac{A_{\rm H}}{4} \ .
\label{entro} \end{equation}

The surface gravity of the black hole solutions is obtained from \cite{wald}
\begin{equation}
\kappa_{\rm sg}^2 =
 - \frac{1}{2} (\nabla_\mu \chi_\nu)(\nabla^\mu \chi^\nu)
\ , \label{sgwald} \end{equation}
with Killing vector
$\chi$, Eq.~(\ref{chi}).
Inserting the expansion at the horizon,
Eqs.~(\ref{H4_hor_red}),
yields the dimensionless surface gravity
\begin{equation}
\kappa_{\rm sg} = \frac{f_2(\theta)}{x_{\rm H} \sqrt{m_2(\theta)}}
\ . \label{temp} \end{equation}
As seen from Eq.~(\ref{relation_hor_1}),
$\kappa_{\rm sg}$ is indeed constant on the horizon,
as required by the zeroth law of black hole mechanics.
The dimensionless temperature $T$ 
of the black hole is proportional to the surface gravity,
\begin{equation}
T = \frac{\kappa_{\rm sg}}{2 \pi} 
\ . \label{tempt} \end{equation}

\subsection{\bf Electric and magnetic charge}

A gauge-invariant definition of
the electromagnetic field strength tensor is
given by the 't Hooft tensor \cite{tHooft}
\begin{equation}
{\cal F}_{\mu\nu} = {\rm Tr} \left\{ \hat \Phi F_{\mu\nu}
- \frac{i}{2e} \hat \Phi D_\mu \hat \Phi D_\nu \hat \Phi \right\}
= \hat \Phi^a F_{\mu\nu}^a + \frac{1}{e} \epsilon_{abc}
\hat \Phi^a D_\mu \hat \Phi^b D_\nu \hat \Phi^c
\ , \label{Hooft_tensor} \end{equation}
where $\hat \Phi$ is the normalized Higgs field,
$|\hat \Phi|^2 = (1/2) {\rm Tr\,} \hat \Phi^2 =\sum_a (\hat \Phi^a)^2 = 1$.

The 't Hooft tensor yields the electric current $j_{\rm el}^\nu$
\begin{equation}
 \nabla_\mu {\cal F}^{\mu\nu} = -4 \pi j_{\rm el}^\nu
\ , \label{jel} \end{equation}
and the magnetic current  $j_{\rm mag}^\nu$
\begin{equation}
 \nabla_\mu {^*}{\cal F}^{\mu\nu} = 4 \pi j_{\rm m}^\nu
\ , \label{jmag} \end{equation}
where ${^*}{\cal F}$ represents the dual field strength tensor.

The electric charge $\cal Q$ is given by
\begin{equation}
 {\cal Q} = \frac{Q}{e} = \frac{1}{4\pi} \int_{S_2}
  {^*}{\cal F}_{\theta\varphi}
 d\theta d\varphi 
\ , \label{Qel} \end{equation}
where the integral is evaluated at spatial infinity.

To define the magnetic charge, we rewrite the  't Hooft tensor as
\begin{equation}
{\cal F}_{\mu\nu} = \partial_\mu{\cal A}_\nu - \partial_\nu{\cal A}_\mu
           -\frac{i}{2 e}
 {\rm Tr}\left\{\hat \Phi \partial_\mu \hat \Phi \partial_\nu \hat \Phi \right\}
\ , \label{Hooft_tensor2} 
\end{equation}
with ${\cal A}_\mu = {\rm Tr}\left\{\hat \Phi A_\mu\right\}$.
Now it follows from Eq.~(\ref{jmag}) that the  magnetic current
$j_{\rm m}^\sigma$ and the topological current $k^\sigma$ are related by 
\begin{equation}
j_{\rm m}^\sigma = \frac{i}{16\pi e}
\epsilon^{\sigma\rho\mu\nu}
{\rm Tr}\left\{
\partial_\rho\hat \Phi \partial_\mu \hat \Phi \partial_\nu \hat \Phi 
\right\} = \frac{1}{e} k^\sigma
\ . \label{jmag2} 
\end{equation}
For globally regular solutions the integration of the magnetic charge density
reduces to a surface integral at spacial infinity which yields
$$
{\cal P} = \frac{n}{e}\varepsilon  \ .
$$
For black hole solutions
we define the magnetic charge by its value on the horizon plus a volume integral,
\begin{equation}
{\cal P} = {\cal P}_{\rm H} + \int_\Sigma (-j_{{\rm m} \, \mu} n^\mu ) dV
  =  {\cal P}_{\rm H} +\int_{r_{\rm H}}^\infty j_{\rm m}^0 \sqrt{-g} dr d\theta d\varphi
\ , \label{Pmag}
\end{equation}
where $\Sigma$ now denotes an asymptotically flat spacelike hypersurface 
bounded by the horizon ${\rm H}$, $dV$ is the natural volume element on 
$\Sigma$, and $n^\mu$ is normal to $\Sigma$ with $n_\mu n^\mu = -1$.

In order to define the horizon magnetic charge we consider the normalized Higgs field
at the horizon as a map between two two-dimensional spheres, which can be 
characterised by a topological number, 
\begin{equation}
N_{\rm H} = \frac{-i}{16 \pi} 
\int_{\rm H} {\rm Tr}\left\{\hat \Phi d\hat \Phi \wedge  d\hat \Phi\right\} \ ,
\end{equation}
and obtain $ {\cal P}_{\rm H} = N_{\rm H}/e$.
For the evaluation of the magnetic charge we note that the volume integral
reduces to a surface integral. Its contribution from the horizon cancels exactly
the horizon magnetic charge, and the contribution from the asymptotic region 
yields 
${\cal P} = \varepsilon n/e$.
Note that for odd $m$ the horizon magnetic charge is either equal to the
magnetic charge or to its negative value, depending on how often the Higgs field function 
$\Phi_1$ 
changes sign on the symmetry axis. For even $m$ both the magnetic charge 
and the horizon magnetic charge are zero.

\subsection{\bf Physical Interpretation of $\nu$ }
The quantity $\nu$ is related to the asymptotic behaviour of the 
gauge potential $A_0$, and therefore it is not defined in a 
gauge-invariant way. To find a physical interpretation of $\nu$ we
apply a gauge transformation that leads to an asymptotically trivial 
gauge potential (for even $m$). Such a gauge transformation is given by
$$
U = e^{i \nu t \tau_z/2} \ e^{ i m \theta \tau_\varphi^{(n)}/2} \ .
$$
The transformed gauge potential and Higgs field are found to be 
\begin{eqnarray}
A_\mu dx^\mu
 & = &  \left( \left[\bar B_1 -\nu-n\frac{\omega}{x}(\cos(m\theta)-1)\right]\frac{\tau_z}{2e} 
  +\left[ B_2 +n\frac{\omega}{x}\sin(m\theta)\right]\frac{\tau_\rho^{(n,\nu t)}}{2e}
 \right) dt 
\nonumber \\
& & 
+A_\varphi (d\varphi-\frac{\omega}{x} dt)
+\left(\frac{H_1}{x}dx +(1-H_2-m)d\theta \right)\frac{\tau_\varphi^{(n,\nu t)}}{2e}
 , \label{gt_a1} 
 \end{eqnarray}
with
\begin{equation}
A_\varphi=   -n\sin\theta\left(\left[H_3 +\frac{\cos(m\theta)-1}{\sin\theta}\right]
\frac{\tau_z}{2e}
   +\left[1-H_4-\frac{\sin(m\theta)}{\sin\theta}\right] \frac{\tau_\rho^{(n,\nu t)}}{2e}\right)  
 , \label{gt_a2} \end{equation}
and 
\begin{equation}
\Phi = \left( \Phi_1 \tau_z + \Phi_2 \tau_\rho^{(n,\nu t)} \right)
\ , \label{gt_a4} \end{equation}
respectively,
where now 
\begin{eqnarray}
\tau_\rho^{(n,\nu t)}  & = & \cos(n \varphi -\nu t) \tau_x + \sin(n \varphi -\nu t) \tau_y \ , 
\nonumber\\
\tau_\varphi^{(n,\nu t)} & = & -\sin(n \varphi -\nu t) \tau_x + \cos(n \varphi -\nu t) \tau_y \ .
\nonumber
\end{eqnarray}
We observe that in this gauge the fields are explicitly time dependent and rotate 
in internal space about the $\tau_z$ direction. The quantity $\nu$ is exactly  
the rotation frequency.

In the presence of magnetic charge, i.e. odd $m$, the transformed gauge potential 
is singular on the negative $z$ axis. However, the physical interpretation of $\nu$ does not
change.

\section{\bf Stationary globally regular EYMHD solutions}

\subsection{\bf Global Charges}

\noindent {\sl \bf  Mass, angular momentum and dilaton charge}

We begin by recalling the general expressions \cite{wald}
for the global mass
\begin{equation}
{\cal M} = 
  \frac{1}{{4\pi G}} \int_{\Sigma}
 R_{\mu\nu}n^\mu\xi^\nu dV
\ , \label{r-relmassy}
 \end{equation}
and the global angular momentum
\begin{equation}
{\cal J} =  
 -\frac{1}{{8\pi G}} \int_{\Sigma}
 R_{\mu\nu}n^\mu\eta^\nu dV
\ . \label{r-relangmom}
\end{equation}
Here $\Sigma$ denotes an asymptotically flat spacelike hypersurface,
$n^\mu$ is normal to $\Sigma$ with $n_\mu n^\mu = -1$,
and $dV$ is the natural volume element on $\Sigma$ \cite{wald}.

Now we express the Ricci tensor in terms of the Yang-Mills, Higgs 
and dilaton fields, using the Einstein equations, the definition of the
stress energy tensor and the Lagrangian
\begin{eqnarray}
\frac{1}{8 \pi G} R_{\mu\nu} & = &
\partial_\mu \Psi \partial_\nu \Psi
+ 2 e^{2\kappa\Psi} {\rm Tr} ( {F_\mu}^\alpha F_{\nu\alpha} )
-\frac{1}{2}e^{2\kappa\Psi} {\rm Tr} ( F_{\rho\sigma} F^{\rho\sigma} )g_{\mu\nu} 
\nonumber \\
& & 
+ \frac{1}{2} {\rm Tr} (D_\mu \Phi D_\nu \Phi) +
\frac{\lambda}{8} e^{-2\kappa \Psi} {\rm Tr} (\Phi^2 -v^2)^2 g_{\mu \nu}
\ . \label{r-mass3}
\end{eqnarray}
Next we replace the third and the last term in Eq.~(\ref{r-mass3})
via the dilaton equation
\begin{equation}
\frac{1}{8 \pi G} R_{\mu\nu} =
\partial_\mu \Psi \partial_\nu \Psi
+ 2 e^{2\kappa\Psi} {\rm Tr} ( {F_\mu}^\alpha F_{\nu\alpha} )
+ \frac{1}{2} {\rm Tr} (D_\mu \Phi D_\nu \Phi) 
-\frac{1}{2\kappa}\frac{1}{\sqrt{-g}}
\partial_\lambda(\sqrt{-g}\partial^\lambda\Psi) g_{\mu\nu} \ .
\label{r-mass4}
\end{equation}
Since $\xi$ and $\eta$ are Killing vector fields and
since $\eta$ is tangential to $\Sigma$, we have
\begin{equation}
\xi^\mu \partial_\mu \Psi = 0\ , \ \ \
\eta^\mu \partial_\mu \Psi = 0\ , \ \ \
n^\mu\eta^\nu g_{\mu\nu} = 0 \ ,
\label{r-mass3a}
\end{equation}
and consequently,
\begin{eqnarray}
\frac{1}{8 \pi G} R_{\mu\nu} n^\mu\xi^\nu
& = &
2 e^{2\kappa\Psi} {\rm Tr} ( {F_\mu}^\alpha F_{\nu\alpha} )n^\mu\xi^\nu
+ \frac{1}{2} {\rm Tr} (D_\mu \Phi D_\nu \Phi)n^\mu\xi^\nu
-\frac{1}{2\kappa}\frac{1}{\sqrt{-g}}
\partial_\lambda(\sqrt{-g}\partial^\lambda\Psi) n^\mu\xi_\mu  \ ,
\label{r-mass3b} \\
\frac{1}{8 \pi G} R_{\mu\nu} n^\mu\eta^\nu
& = &
2 e^{2\kappa\Psi} {\rm Tr} ( {F_\mu}^\alpha F_{\nu\alpha} )n^\mu\eta^\nu +
\frac{1}{2} {\rm Tr} (D_\mu \Phi D_\nu \Phi)n^\mu\eta^\nu\  \ .
\label{r-mass3c}
\end{eqnarray}

We now define the dilaton charge $\cal D$ via
\begin{equation}
\int_{\Sigma}
\frac{1}{\sqrt{-g}}
\partial_\lambda(\sqrt{-g}\partial^\lambda\Psi) n^\mu\xi_\mu\ dV
= - 4 \pi {\cal D} \ .
\label{dil-charge}
\end{equation}
Making use of the dilaton charge $\cal D$, we obtain for the mass $\cal M$
\begin{equation} \hspace{1cm}
{\cal M} =
4 \int_\Sigma\left\{
        e^{2\kappa\Psi}{\rm Tr} ({F_\mu}^\alpha F_{\nu\alpha})n^\mu\xi^\nu
\right\} dV 
+\int_\Sigma\left\{{\rm Tr} (D_\mu\Phi D_\nu \Phi)n^\mu\xi^\nu
\right\} dV  
+ \frac{4\pi}{\kappa} {\cal D}
\ , \label{r-mass5}
\end{equation}
while the angular momentum $\cal J$ is given by
\begin{equation}
{\cal J} =
- 2 \int_\Sigma\left\{
        e^{2\kappa\Psi}{\rm Tr} ({F_\mu}^\alpha F_{\nu\alpha})n^\mu\eta^\nu
\right\} dV 
- \frac{1}{2}\int_\Sigma\left\{{\rm Tr} (D_\mu\Phi D_\nu \Phi)n^\mu\eta^\nu
\right\} dV  
\ . \label{r-mass5a}
\end{equation}

To evaluate the integrals in Eq.~(\ref{r-mass5}) and
Eq.~(\ref{r-mass5a}) we use local coordinates
$(t,r,\theta, \varphi)$. In these coordinates
\begin{equation}
n^\mu = -\sqrt{f} g^{0\mu} \ , \ \ \
\xi^\mu =(1,0,0,0)\ , \ \ \
\eta^\mu =(0,0,0,1)\ , \ \ \
dV=\frac{1}{\sqrt{f}}\sqrt{-g}\, dr d\theta d\varphi \ ,
\label{r-mass5b}
\end{equation}
and we obtain
\begin{equation} \hspace{-1cm}
{\DS {\cal M} - \frac{4\pi}{\kappa} {\cal D}
 = {\cal I_M}} \equiv {\DS - 4 \int_\Sigma  e^{2 \kappa \Psi} {\rm Tr}
 \left[   F_{0\mu} F^{0\mu} \right]
 \sqrt{-g} dr d\theta d\varphi} 
 {\DS -  \int_\Sigma  {\rm Tr}\left[ D_0 \Phi D^0\Phi \right]
 \sqrt{-g} dr d\theta d\varphi} \ ,
\label{r-intIM}
\end{equation}
\begin{equation}
{\DS {\cal J}
 = {\cal I_J}} \equiv {\DS  2 \int_\Sigma  e^{2 \kappa \Psi} {\rm Tr}
 \left[  F_{\varphi\mu} F^{0\mu} \right]
 \sqrt{-g} dr d\theta d\varphi} 
 {\DS + \frac{1}{2}
  \int_\Sigma  {\rm Tr}\left[  D_{\varphi}\Phi D^0\Phi \right]
 \sqrt{-g} dr d\theta d\varphi} \ ,
\label{r-intIJ}
\end{equation}
defining the integrals ${\cal I_M}$ and ${\cal I_J}$.

To evaluate the integrals ${\cal I_M}$, Eq.~(\ref{r-intIM})
and ${\cal I_J}$, Eq.~(\ref{r-intIJ}), we make use of the
symmetry relations, Eqs.~(\ref{symA}) \cite{eugen},
\begin{equation}
F_{\mu 0} = \hD_\mu A_0 \ , \ \ \
F_{\mu \varphi} = \hD_\mu \left( A_\varphi- W_\eta \right)
\ , \label{r-fsymA} \end{equation}
where $\hD_\mu \equiv \partial_\mu + ie[A_\mu,\cdot \ ]$.
The integrals then read
\begin{equation}
 {\cal I_M} =
{\DS + 4 \int_\Sigma  e^{2 \kappa \Psi} {\rm Tr}
 \left[   \hD_\mu A_0 F^{0\mu} \right]
 \sqrt{-g} dr d\theta d\varphi}
 {\DS -  \int_\Sigma  {\rm Tr}\left[ D_0 \Phi D^0\Phi \right]
 \sqrt{-g} dr d\theta d\varphi} 
\ , \label{r-mass6a} \end{equation}
\begin{equation}
 {\cal I_J} =
- {\DS  2 \int_\Sigma  e^{2 \kappa \Psi} {\rm Tr}
 \left[  \hD_\mu \left( A_{\varphi} - W_\eta \right) F^{0\mu} \right]
 \sqrt{-g} dr d\theta d\varphi}
 {\DS + \frac{1}{2}
  \int_\Sigma  {\rm Tr}\left[  D_{\varphi}\Phi D^0\Phi \right]
 \sqrt{-g} dr d\theta d\varphi} 
\ . \label{r-mass6b} \end{equation}

Adding zero to the above integrals, in the form of the
gauge field equation of motion for the zero component,
we obtain
\begin{equation}
 {\cal I_M} =
  \displaystyle 4 \int_\Sigma
 {\rm Tr}
   \left[ \hD_\mu \left\{ 
    A_0  e^{2 \kappa \Psi} F^{0\mu} \sqrt{-g} \right\} \right]
   dr d\theta d\varphi 
 + \int_\Sigma
 {\rm Tr}
   \left[ ie
    A_0 [\Phi,D^0\Phi] - D_0 \Phi D^0\Phi
 \right] \sqrt{-g} dr d\theta d\varphi 
\ , \label{IM}
\end{equation}
\begin{eqnarray}
 {\cal I_J}  & = &
  \displaystyle -2 \int_\Sigma
 {\rm Tr}
   \left[ \hD_\mu \left\{
    \left( A_{\varphi} - W_\eta \right)
  e^{2 \kappa \Psi} F^{0\mu} \sqrt{-g} \right\} \right]
   dr d\theta d\varphi
\nonumber \\
& &
 - \frac{1}{2} \int_\Sigma
 {\rm Tr}
   \left[ ie
    \left( A_{\varphi} - W_\eta \right) [\Phi,D^0\Phi] - D_\varphi \Phi D^0\Phi
 \right] \sqrt{-g} dr d\theta d\varphi
\ . \label{IJ}
\end{eqnarray}
Making use of the explicit form of the ansatz,
exploiting in particular Eq.~(\ref{symh}),
we see, that for both ${\cal I}_M$ and ${\cal I}_J$ 
the second integral vanishes identically,
leaving only the first integral to be analyzed further.

Since the trace of a commutator vanishes,
we now replace the derivative $\hD_\mu$ 
by the partial derivative $\partial_\mu$ in the remaining integrals,
\begin{equation}
 {\cal }I_M =
  {\displaystyle 4 \int_\Sigma
 {\rm Tr}  \left[ \partial_\mu \left\{ 
    A_0 
   e^{2 \kappa \Psi} F^{0\mu} \sqrt{-g} \right\} \right]
   dr d\theta d\varphi}
\ , \label{IIa} \end{equation}
\begin{equation}
 {\cal I}_J =
  {\displaystyle - 2 \int_\Sigma
 {\rm Tr}  \left[ \partial_\mu \left\{ 
    \left( A_{\varphi} - W_\eta \right)
    e^{2 \kappa \Psi} F^{0\mu} \sqrt{-g} \right\} \right]
   dr d\theta d\varphi}
\ , \label{IIb} \end{equation}
and employ the divergence theorem.
The $\theta$-term vanishes,
since $\sqrt{-g}$ vanishes at $\theta=0$ and $\theta=\pi$,
and the $\varphi$-term vanishes,
since the integrands at $\varphi=0$ and $\varphi=2\pi$ coincide,
thus we are left with
\begin{equation}
{\cal I}_M=
 {\displaystyle 4 \int \left.
 {\rm Tr}  \left[  A_0 
 e^{2 \kappa \Psi} F^{0r} \sqrt{-g} \right]
 \right|^\infty_{0} d\theta d\varphi}
\ , \label{intM} \end{equation}
\begin{equation}
{\cal I}_J=
 {\displaystyle -2 \int \left.
 {\rm Tr}  \left[  
   \left( A_{\varphi} - W_\eta \right)  
 e^{2 \kappa \Psi} F^{0r} \sqrt{-g} \right]
 \right|^\infty_{0} d\theta d\varphi}
\ . \label{intJ} \end{equation}

Since the integrands vanish at the origin,
the only contributions to ${\cal I}_M$ 
and ${\cal I}_J$ come from infinity.
At infinity the asymptotic expansion yields
to lowest order
\begin{eqnarray}
F^{0r} \sqrt{-g} & = &  Q \sin \theta \frac{\tau_r^{(n,m)}}{2e}
 + o(1)
\ , \nonumber \\
A_0 & = & {\tilde \nu} \frac{\tau_r^{(n,m)}}{2e} + o(1)
\ , \nonumber \\
A_\varphi & = &  -n\sin\theta \left[ \frac{\cos(\varepsilon\theta)-\cos(m\theta)}{\sin\theta}\frac{\tau_r^{(n,m)}}{2e}
+ \frac{\sin(m\theta)}{\sin\theta} \frac{\tau_\theta^{(n,m)}}{2e} \right]  + o(1)
\ , \label{r-A_asym} \end{eqnarray}
where
\begin{equation}
{\tilde \nu}=\frac{e\alpha}{\sqrt{4\pi G}} \nu \ , \ \ \ \varepsilon=\frac{1}{2}(1-(-1)^m) \ .
\end{equation}
The integrals ${\cal I}_M$ and ${\cal I}_J$ are then given by
\begin{equation}
{\cal I}_M =  \frac{8\pi{\tilde \nu} Q}{e^2}  \ , \ \ \
{\cal I}_J =
  \frac{4\pi n Q}{e^2} (1-\varepsilon) 
\ , \label{r-IIfinal} \end{equation}
yielding for the mass $\cal M$ and the angular momentum $\cal J$
\begin{equation}
{\cal M} = \frac{4\pi}{\kappa} {\cal D} + \frac{8\pi{\tilde \nu} Q}{e^2} 
 \ , \ \ \ 
{\cal J} =  \frac{4\pi  n Q}{e^2} (1-\varepsilon)
\ . \label{r-Ifinal} \end{equation}

Returning to dimensionless variables, and noting that
\begin{equation}
{\cal M}= \frac{\sqrt{4\pi G}}{e\alpha G}\mu \ , \ \ \
{\cal J} =\frac{4\pi}{e^2\alpha^2}\zeta \ , \ \ \
{\cal D} = \frac{D}{e} \ ,
\label{dimlessMJD}
\end{equation}
we obtain the mass formula
\begin{equation}
\mu = \alpha^2 \frac{D}{\gamma} +2\alpha^2\nu Q 
\Longleftrightarrow M = \frac{D}{\gamma} +2\nu Q \ ,
\label{Mreg}
\end{equation}
and the quantization condition for the angular momentum Eq.~(\ref{quant})
\begin{equation}
\zeta= \alpha^2 n Q (1-\varepsilon)
\Longleftrightarrow 
J= n Q (1-\varepsilon)
\ . \nonumber
\end{equation}

\subsection{\bf Effective Action}

To address the dependence of the globally regular solutions 
on the coupling constant $\alpha$, we now consider the effective action
${S}^{\rm eff}$.
In particular, we explain the qualitatively
different dependence of the mass $M$ for static and for stationary solutions.
This concerns only such types of regular solutions 
where two branches of solutions exist.

For static solutions, the mass $M$ exhibits a ``spike''
at the maximal value of the coupling $\alpha_{\rm max}$,
where the branches merge and end \cite{gmono,hkk,KKS}.
The tangent of the mass w.r.t.~$\alpha$ must be the same
for both branches at $\alpha_{\rm max}$ \cite{peter}.
In contrast, for stationary solutions,
the mass $M$ exhibits a ``loop''
in the vicinity of the maximal value of the coupling $\alpha_{\rm max}$
\cite{ulrike}.
Here the tangent of the mass w.r.t.~$\alpha$ diverges at $\alpha_{\rm max}$.
The loop is associated with a critical value of $\alpha$,
where the two mass branches cross.
\\

\noindent {\sl \bf Effective action and mass}

Let us begin by defining the effective action
${\cal S}^{\rm eff}$,
\begin{equation}
{\cal S}^{\rm eff}
 =\int \left ( \frac{\hat R}{16\pi G} 
 + {L}_M \right ) \sqrt{-g} d^3x \ ,
\ \label{action-eff} \end{equation}
with the gravitational effective Lagrangian 
\begin{equation}
\frac{\hat R}{16\pi G} = \frac{1}{16\pi G} \left(
 R - \frac{\partial_\mu \Delta^\mu}{\sqrt{-g}} \right)
\ \label{R-eff} \end{equation}
and the matter Lagrangian ${L}_M$ Eq.~(\ref{lagm})
$$
{L}_M=-\frac{1}{2}\partial_\mu \Psi \partial^\mu \Psi
 - \frac{1}{2} e^{2 \kappa \Psi } {\rm Tr} (F_{\mu\nu} F^{\mu\nu})
-\frac{1}{4} {\rm Tr} \left( D_\mu \Phi D^\mu \Phi \right)
-\frac{\lambda}{8} e^{-2 \kappa \Psi } {\rm Tr}
 \left( \Phi^2 - v^2 \right)^2 \ .
$$
The divergence term $\partial_\mu \Delta^\mu$ 
in the gravitational effective Lagrangian ensures
that the varational principle of the effective action ${\cal S}^{\rm eff}$
w.r.t.~the functions of the ansatz yields the proper set of
field equations.
For our particular ansatz of the metric
$\Delta^\mu$ is given by
\begin{equation}
\Delta^\mu = \sqrt{l} \sin \theta \
 \left( 0\, , r^2 \partial_r\, , \partial_\theta\, , 0\, \right)
\, \ln \frac{f}{ml}
\ . \label{Delta-mu}
\end{equation}

Reexpressing the curvature scalar $R$ via the Einstein equations,
\begin{equation}
\frac{R}{8 \pi G} = 
\partial_\mu \Psi \partial^\mu \Psi
+\frac{1}{2} {\rm Tr} \left( D_\mu \Phi D^\mu \Phi \right)
+\frac{\lambda}{2} e^{-2 \kappa \Psi } {\rm Tr}
 \left( \Phi^2 - v^2 \right)^2
\ , \label{Rex} \end{equation}
then leads to the effective action
\begin{equation}
{\cal S}^{\rm eff}
 =\int \left( - \frac{1}{16 \pi G}
\, \frac{\partial_\mu \Delta^\mu}{\sqrt{-g}}
- \frac{1}{2} e^{2 \kappa \Psi } {\rm Tr} (F_{\mu\nu} F^{\mu\nu})
+\frac{\lambda}{8} e^{-2 \kappa \Psi } {\rm Tr}
 \left( \Phi^2 - v^2 \right)^2
\right) \sqrt{-g} d^3 x
\ . \label{Seff-2} \end{equation}

Analogously to the derivation of the mass formula, we next
replace the two matter terms in ${\cal S}^{\rm eff}$
via the equation of motion of the dilaton field and obtain
\begin{equation}
{\cal S}^{\rm eff}
 = - \frac{1}{16 \pi G} \int \partial_\mu \Delta^\mu dr d \theta d\varphi
 +\frac{1}{2\kappa} \int \frac{1}{\sqrt{-g}}
\partial_\lambda(\sqrt{-g}\partial^\lambda\Psi) \sqrt{-g} dr d \theta d\varphi
\ . \label{Seff-3} \end{equation}
Since in the local coordinates the second integral agrees with the integral
for the dilaton charge Eq.~(\ref{dil-charge}) and since the $\theta$-term
in the first integral vanishes, the effective action becomes
\begin{equation}
{\cal S}^{\rm eff}
= - \frac{1}{16 \pi G} \int \Delta^r d \theta d \varphi
 - \frac{4 \pi {\cal D}}{2 \kappa}
\ . \label{Seff-4} \end{equation}
The remaining integral is evaluated with help of
the asymptotic expansion of the metric functions,
leading to
\begin{equation}
{\cal S}^{\rm eff} = - \frac{1}{2} \left(
 {\cal M} + \frac{4 \pi {\cal D}}{ \kappa} \right)
\ , \label{Seff-5} \end{equation}
which can be rewritten via the mass formula
for the regular solutions Eq.~(\ref{r-Ifinal})
\begin{equation}
{\cal S}^{\rm eff} = - {\cal M} + \frac{4\pi{\tilde \nu} Q}{e^2}
\ . \label{Seff-6} \end{equation}

Defining finally the dimensionless effective action
$S^{\rm eff}$
\begin{equation}
{\cal S}^{\rm eff} = 
                     \frac{\sqrt{4\pi G}}{e\alpha G} \, \alpha^2 S^{\rm eff}
\ , \label{Seff-dimless} \end{equation}
we obtain
\begin{equation}
S^{\rm eff} = -\left( M -\nu Q \right)
\   \label{Seff-7} \end{equation}
or equivalently
\begin{equation}
S^{\rm eff} = -  \left(\frac{D}{\gamma} + \nu Q\right)
\ . \label{Seff-8} \end{equation}
Note, that Eq.~(\ref{Seff-7}) for the effective action
remains true when the dilaton decouples,
i.e.~Eq.~(\ref{Seff-7}) also holds for EYMH solutions.\\

\noindent {\sl \bf 
Dependence of the effective action $S^{\rm{eff}}$ and the mass $M$ on 
the coupling constant $\alpha$}

To address the dependence of the effective action $S^{\rm{eff}}$ on
the coupling constant $\alpha$
we first make the $\alpha$-dependence more explicit.
To this end, we express all quantities in the effective action
in dimensionless quantities.
Denoting the by the divergence term corrected 
dimensionless curvature scalar $\bar R$ (i.e.~$\hat R \rightarrow \bar R$),
the dimensionless matter Lagrangian $\bar L_M$,
and the dimensionless determinant of the metric $-\bar g$, and 
we obtain the dimensionless effective action, 
\begin{equation}
S^{\rm eff}= \frac{1}{4 \pi}
\left[ \frac{1}{4\alpha^2} \int \bar R \sqrt{-\bar g} d^3x
 + \int \bar L_M \sqrt{-\bar g} d^3x  \right]
\ . \label{arg1} \end{equation}

We now take the derivative of $S^{\rm eff}$
w.r.t.~$\alpha$, taking into account
that the metric and matter functions, abbreviated by $X_i$,
implicitly also depend on $\alpha$. The derivative
has thus two terms
\begin{equation}
\frac{ d S^{\rm{eff}} }{d \alpha} =
\frac{1}{4 \pi}\left[
-\frac{1}{2\alpha^3} \int\bar R \sqrt{-\bar g}   d^3x
\right]
+
\frac{1}{16 \pi \alpha^2} 
 \int\underbrace{ \left\{
 \frac{\partial ( \bar R \sqrt{-\bar g} ) }{\partial X_i}
+4 \alpha^2 \frac{\partial ( \bar L_M \sqrt{-\bar g} ) }
  {\partial X_i}\right\}}_{\mbox{$=0\;\;\; $equations of motion}}
 \frac{\partial X_i}{\partial \alpha} d^3x
\ , \label{arg2} \end{equation}
where the second term vanishes for solutions
of the equations of motion,
and we are left with
\begin{equation}
\frac{ d S^{\rm{eff}} }{d \alpha} =
-\frac{1}{8\pi\alpha^3} \int \bar R \sqrt{-\bar g} d^3x
\ . \label{arg3} \end{equation}

From Eq.~(\ref{arg3}) we conclude, that the effective action $S^{\rm{eff}}$
must exhibit a ``spike''
at the maximal value of the coupling $\alpha_{\rm max}$,
where the two branches of solutions merge and end,
since the tangent w.r.t.~$\alpha$ must be the same
for both branches at $\alpha_{\rm max}$.
This conclusion holds for stationary solutions,
as well as for static solutions.

Let us now address the mass $M$, which is related to the effective
action via Eq.~(\ref{Seff-7}), i.e.
\begin{equation}
M = - S^{\rm eff} + \nu Q
\ . \label{arg4} \end{equation}
Considering the derivative of the mass $M$ w.r.t.~$\alpha$,
keeping $\nu$ fixed, we obtain
\begin{equation}
\frac{ d M}{d \alpha} =
- \frac{ d S^{\rm{eff}}}{d \alpha} 
+\nu \frac{ d Q}{d \alpha}
\ . \label{arg5} \end{equation}
Clearly, the crucial difference between the tangent of the mass and 
the tangent of the effective action resides in the last term, containing
the derivative of the electric charge $Q$ w.r.t.~$\alpha$.
It is this term which allows different tangents on both branches
at $\alpha_{\rm max}$. In fact,
for stationary (non-static) solutions, we observe that this last term diverges
at $\alpha_{\rm max}$, yielding a divergent tangent also for the mass,
as required for a ``loop'' associated with 
both mass branches in the vicinity of $\alpha_{\rm max}$.
For static solutions, on the other hand, the mass $M$ 
always exhibits a ``spike''
at the maximal value of the coupling, $\alpha_{\rm max}$,
since the electric charge $Q$ vanishes.

Note however, that we assumed so far that the value of $\nu$ is fixed
by a boundary condition, $ \left. \bar{B}_1\right|_\infty =\nu$.
This is in contrast to the
case where the electric charge is kept fixed by 
the boundary condition at infinity, 
\begin{equation}
\left. x^2 \partial_x \bar{B}_1 \right|_\infty = Q \ .
\end{equation}
In this case the quantity $\nu$
is allowed to vary and its value is adjusted by the numerical 
procedure.
More formally, in the variation of the effective action a boundary term shows
up, which evaluates to $Q \delta \nu$.
Therefore the field equations are obtained from the modified effective 
action, 
\begin{equation}
\delta\left. \left( S^{\rm eff}- Q \nu\right)\right|_Q = 0
\end{equation}
As a consequence, if $\alpha$ is varied for fixed electric charge $Q$, 
\begin{equation}
\frac{d}{d\alpha}\left( \left.S^{\rm eff}- Q \nu\right)\right|_Q
= -\frac{1}{8\pi\alpha^3} \int\bar R \sqrt{-\bar g}   d^3x \ ,
\end{equation}
since the variation of the modified effective action with respect to
the fields vanishes.
On the other hand, since $S^{\rm eff}- Q \nu = -M$,
we find
\begin{equation}
\frac{d M}{d\alpha} 
= \frac{1}{8\pi\alpha^3} \int\bar R \sqrt{-\bar g}  d^3x \ .
\end{equation}
Thus for fixed electric charge it is the mass that exhibits a spike.

\subsection{\bf Numerical results}

We solve the set of thirteen coupled non-linear
elliptic partial differential equations numerically \cite{schoen},
subject to the above boundary conditions, requiring
the solutions to be regular at the origin.
We employ compactified dimensionless coordinates,
$\bar x = x/(1+x)$. 
The numerical calculations, based on the Newton-Raphson method,
are performed with help of the program FIDISOL \cite{schoen}.
The equations are discretized on a non-equidistant grid in 
$\bar x$ and  $\theta$.
Typical grids used have sizes $100 \times 20$, 
covering the integration region
$0\leq \bar x\leq 1$ and $0\leq\theta\leq\pi/2$.
(See \cite{kkreg,kkrot} and \cite{schoen} 
for further details on the numerical procedure.)

For given coupling constants $\alpha$, $\beta$ and $\gamma$,
the stationary globally regular solutions 
then depend on the parameter $\nu$, specifying the
time component of the gauge potential at infinity,
and on the integers $m$ and $n$.
(In principle, the 
solutions can further depend on the node number of the gauge potential
functions $k$, labelling the radial excitations.
However, we here focus on the lowest mass solutions.)

We now illustrate the above relations for the mass 
and the effective action with numerical results presented in Figs.~1 and 2.
We first consider stationary gravitating dyon solutions with $m=1$.
Dyons with magnetic charge $n=1$ are spherically symmetric \cite{gdyon},
dyons with higher magnetic charge are axially symmetric \cite{ulrike}.

In Fig.~1a we exhibit the mass $M$ 
of dyons with magnetic charge $n=1$, 2 and 3 at a fixed value of $\nu$
versus the coupling constant $\alpha$ (at $\beta=\gamma=0$).
In each case, a first branch of gravitating dyons
emerges from the corresponding flat space solution at $\alpha=0$
and extends up to a maximal value of the
coupling constant, $\alpha_{\rm max}$, beyond which no
dyon solutions exist.
For the $n=1$ dyons we observe a second branch of solutions in the vicinity
of the maximal value of $\alpha$.
This second branch ends at a critical value of $\alpha$,
where the branch of non-Abelian solutions merges with 
the corresponding branch of extremal
Reissner-Nordstr\"om solutions \cite{gmono,gdyon}.
For $n>1$ dyon solutions, numerical accuracy does not
allow us to discern the existence of two branches.

\begin{figure}[h!]
\noindent\parbox{\textwidth}{
\centerline{
(a)\mbox{\epsfysize=5.0cm \epsffile{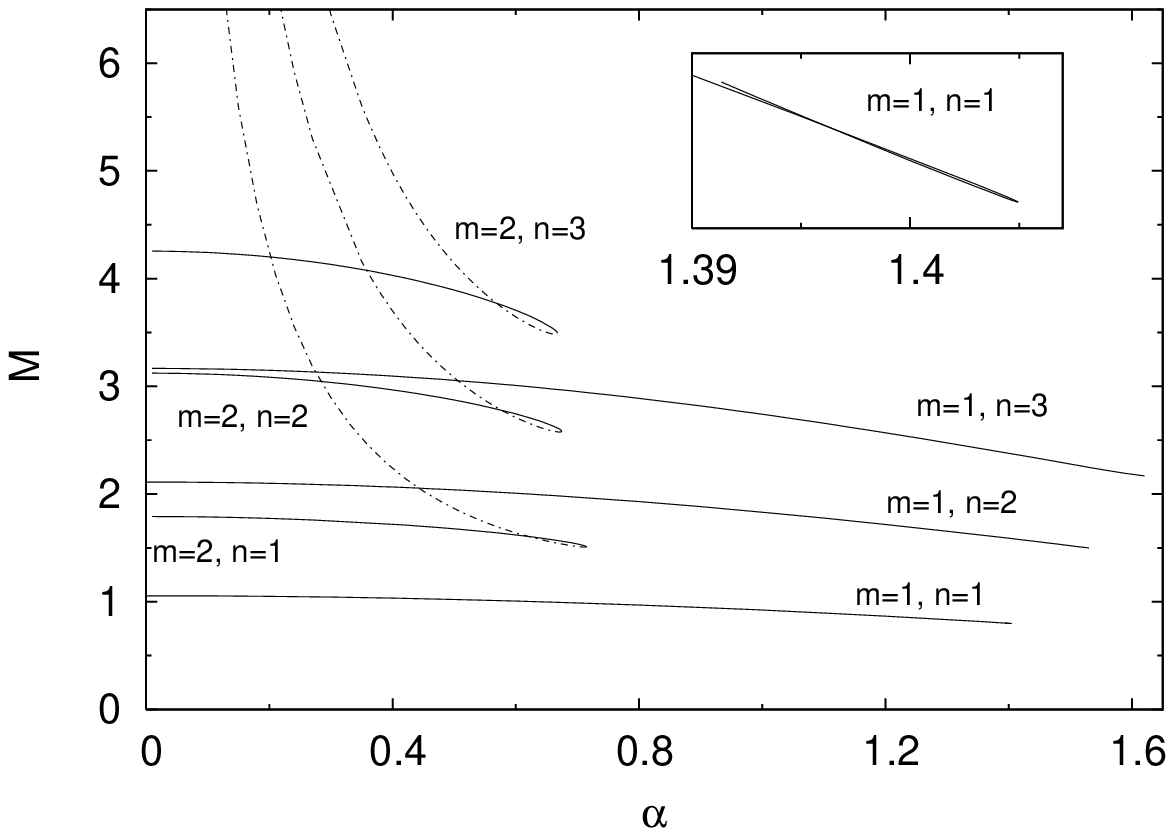} }
(b)\mbox{\epsfysize=5.0cm \epsffile{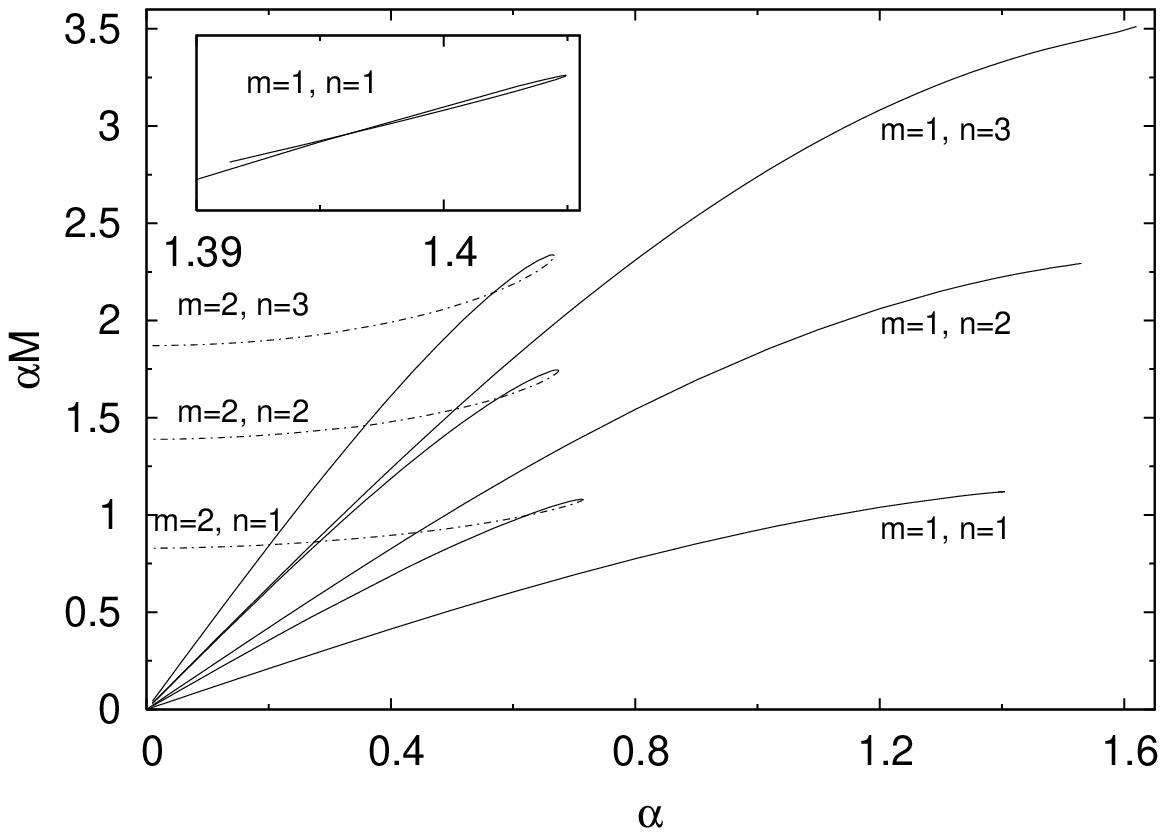} }
}\vspace{0.5cm}
\centerline{
(c)\mbox{\epsfysize=5.0cm \epsffile{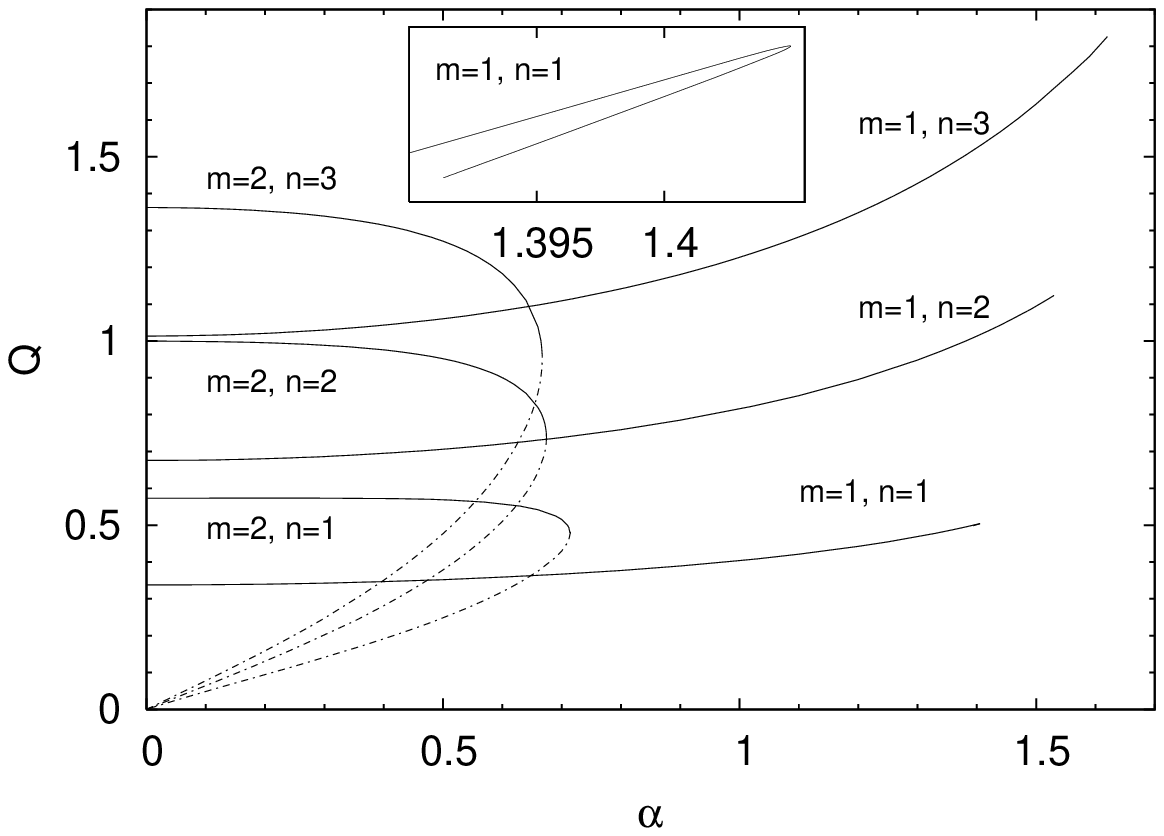} }
(d)\mbox{\epsfysize=5.0cm \epsffile{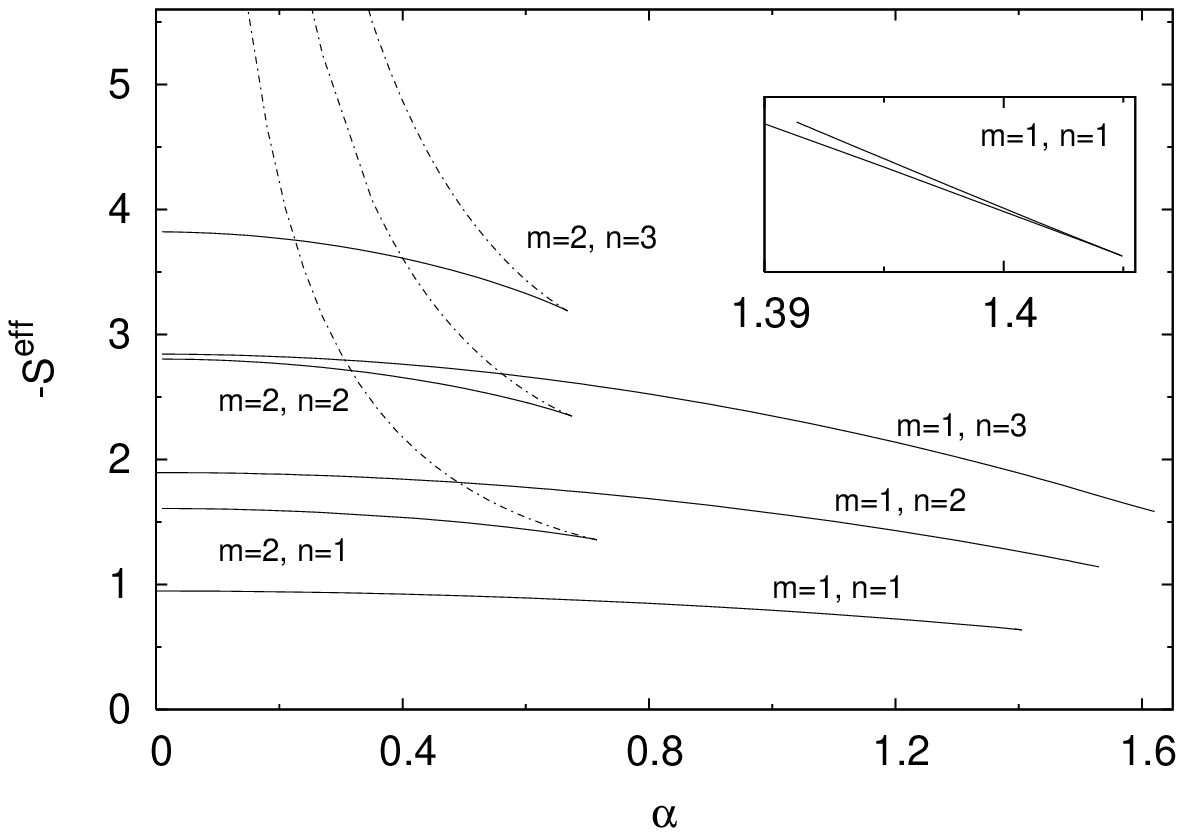} }
}\vspace{0.5cm}
{\bf Fig.~1} \small
The mass $M$ (a), the scaled mass $\alpha M$ (b),
the electric charge $Q$ (c),
and the effective action $S^{\rm eff}$ (d) are
shown versus the coupling constant $\alpha$
for dyon solutions with $m=1$, $n=1,2,3$
and for electrically charged
monopole-antimonopole resp.~vortex ring solutions
with $m=2$, $n=1,2,3$ at $\nu=0.32$, $\beta=\gamma=0$
(first branch: solid, second branch: dot-dashed).
\vspace{0.5cm}
}
\end{figure}

Besides dyons, we also exhibit in Fig.~1a the mass of 
electrically charged monopole-antimonopole pair
resp.~vortex ring solutions,
which have $m=2$ and $n=1$, 2 and 3. 
For these solutions always two branches of solutions exist. 
Again, the first branch emerges from the 
respective flat space solution, 
and extends up to a maximal value of $\alpha$,
where it merges with the second branch.
But the second branch now extends back to $\alpha \rightarrow 0$. 
The mass diverges on the second branch in the limit $\alpha \rightarrow 0$.
But considering the scaled mass $\alpha M$ instead,
exhibited in Fig.~1b, 
one realizes, that in the limit $\alpha \rightarrow 0$
a globally regular EYM solution \cite{bm,kkreg}
is reached (after rescaling) \cite{KKS}.
Clearly, the electric charge $Q$, exhibited in Fig.~1c,
also tends to zero on the second branch in the limit $\alpha \rightarrow 0$,
yielding non-rotating limiting EYM solutions,
in agreement with previous results on globally regular EYM solutions
\cite{bizon,eugen}.

We exhibit the effective action $S^{\rm eff}$ for the same 
set of solutions in Fig.~1d.
As predicted above, the effective action exhibits a ``spike''
close to the maximal value of the coupling constant $\alpha$,
whenever two branches of solutions are present.
The mass, in contrast, exhibits a ``loop'' for these
stationary non-static solutions close to $\alpha_{\rm max}$.
At $\alpha_{\rm max}$ the tangent of the mass diverges, 
since the tangent of the electric charge diverges there.
If we consider branches of static solutions instead,
the mass exhibits a ``spike'' close to $\alpha_{\rm max}$ \cite{gmono,KKS}.

While in the solutions of Fig.~1 the dilaton is decoupled
since $\gamma=0$, we consider in Fig.~2 the dependence of the solutions
on the dilaton coupling constant $\gamma$.
In Fig.~2a we exhibit the mass and the effective action
of electrically charged monopole-antimonopole pair solutions
($m=2$, $n=1$, $\nu=0.32$, $\beta=0$)
versus the coupling constant $\alpha$ at fixed 
dilaton coupling constant $\gamma=0.4$.
As predicted, we observe a ``spike'' for the effective
action and a ``loop'' for the mass.
Note, that the mass of these EYMHD solutions
does not diverge on the second branch in the limit $\alpha \rightarrow 0$,
since a YMHD solution is approached.
We note, that the effect of a dilaton on globally regular solutions
is very similar to the effect of gravity \cite{Forgacs2}.

\begin{figure}[h!]
\noindent\parbox{\textwidth}{
\centerline{
(a)\mbox{\epsfysize=5.0cm \epsffile{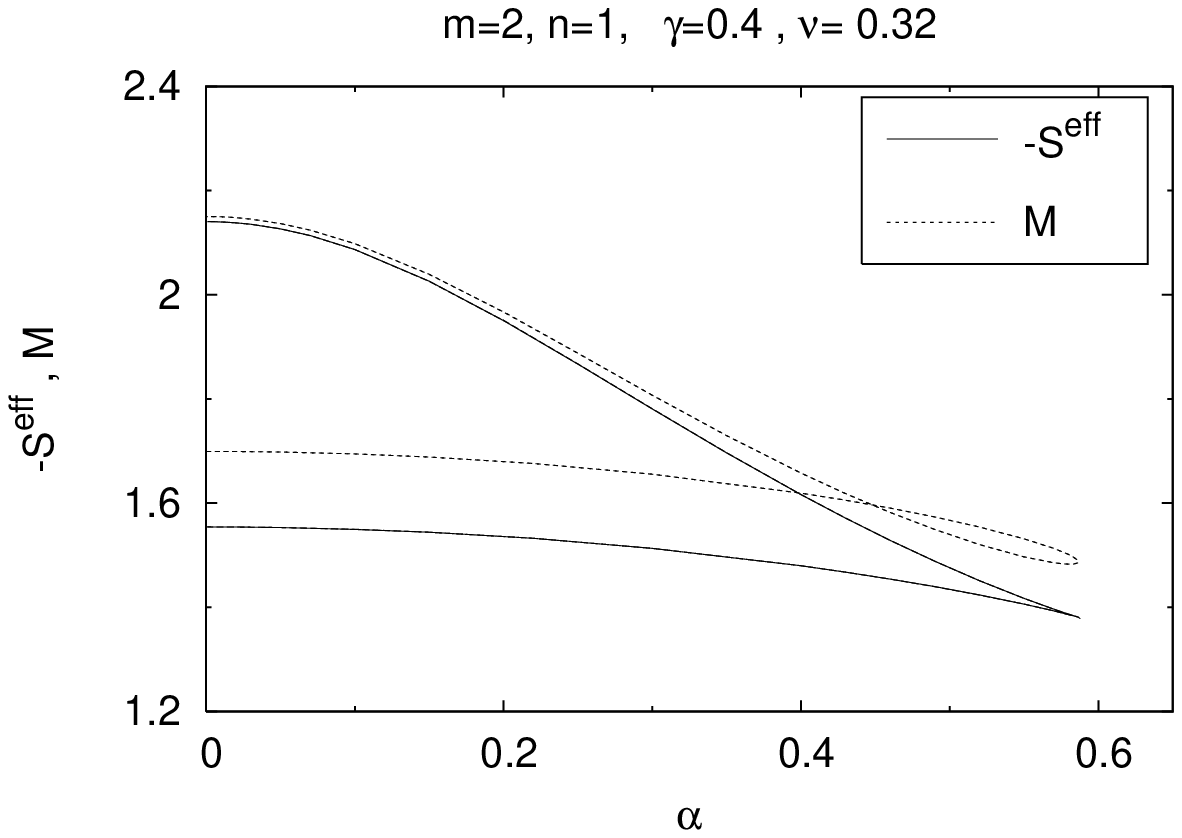} }
(b)\mbox{\epsfysize=5.0cm \epsffile{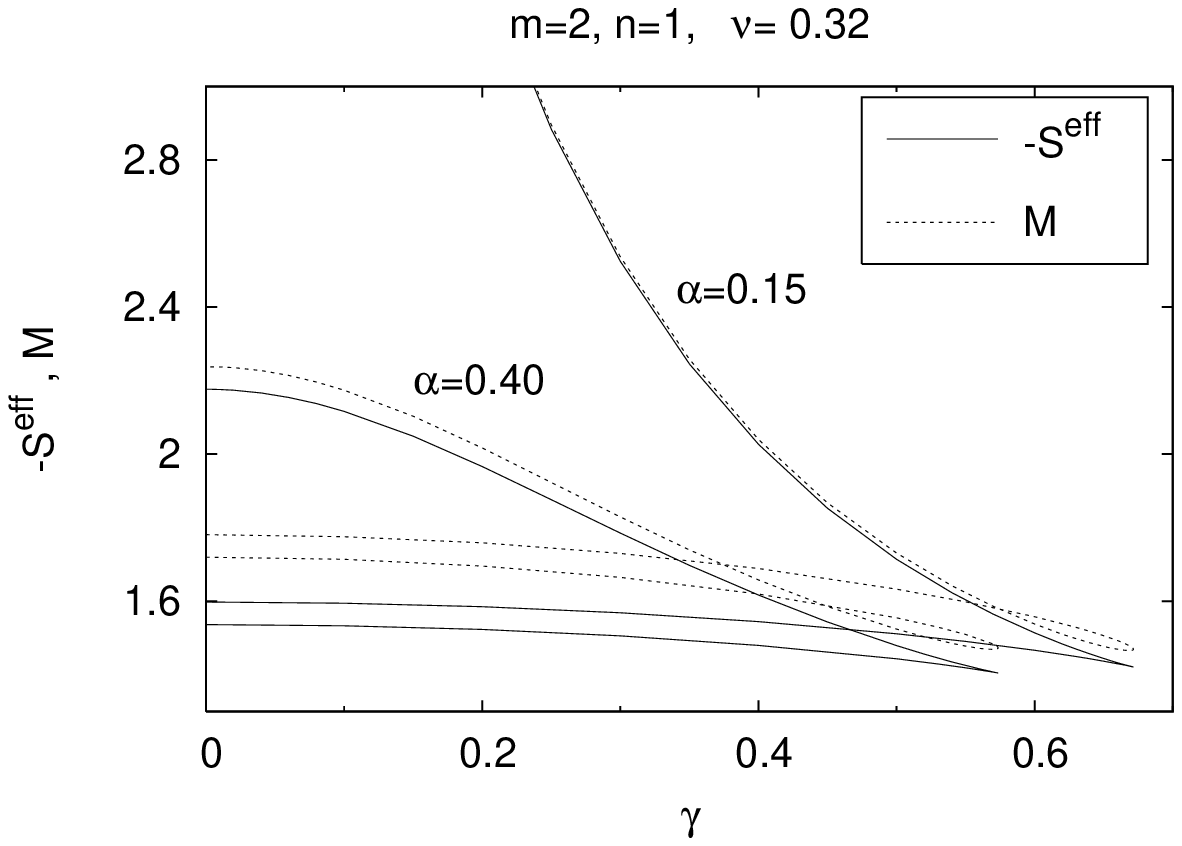} }
}\vspace{0.5cm}
{\bf Fig.~2} \small
The mass $M$ and the effective action $S^{\rm eff}$ 
of electrically charged monopole-antimonopole solutions
with $m=2$, $n=1$, $\nu=0.32$, $\beta=0$ are
shown versus the coupling constant $\alpha$ for
dilaton coupling constant $\gamma=0.4$ (a),
and versus the dilaton coupling constant $\gamma$ for
coupling constant $\alpha=0.15$ and $0.40$ (b).
\vspace{0.5cm}
}
\end{figure}

The $\gamma$-dependence of the solutions (at fixed $\alpha$)
is illustrated in Fig.~2b for
electrically charged monopole-antimonopole pair solutions
with $m=2$, $n=1$, $\nu=0.32$ ($\beta=0$)
for $\alpha=0.15$ and $0.40$.
Again we note, that the effective action exhibits a ``spike''
while the mass exhibits a ``loop''.
That this must be the case,
can be shown by an argument analogous to the one employed above
for the $\alpha$-dependence.
Note, that the mass of these EYMHD solutions
does not diverge on the second branch in the limit $\gamma \rightarrow 0$,
since EYMH solutions are approached in this limit.

We illustrate the globally regular solutions with an example
in Fig.~3.
We here exhibit the energy density of the matter fields $\epsilon$
\begin{equation}
\epsilon = - \frac{2}{e^2 v^4}
 \left( T^0_0 - \frac{1}{2} 
T_\mu^\mu \right)
 \   \label{edens} \end{equation}
for a monopole-antimonopole pair solution ($m=2$, $n=1$) 
carrying electric charge and angular momentum
($\nu=0.32$, $\alpha=0.3$, $\beta=\gamma=0$).
The maxima of the energy density are associated with the
location of the magnetic poles on the symmetry axis.

\begin{figure}[h!]
\noindent\parbox{\textwidth}{
\begin{center}
(a)\epsfysize=9.cm \epsffile{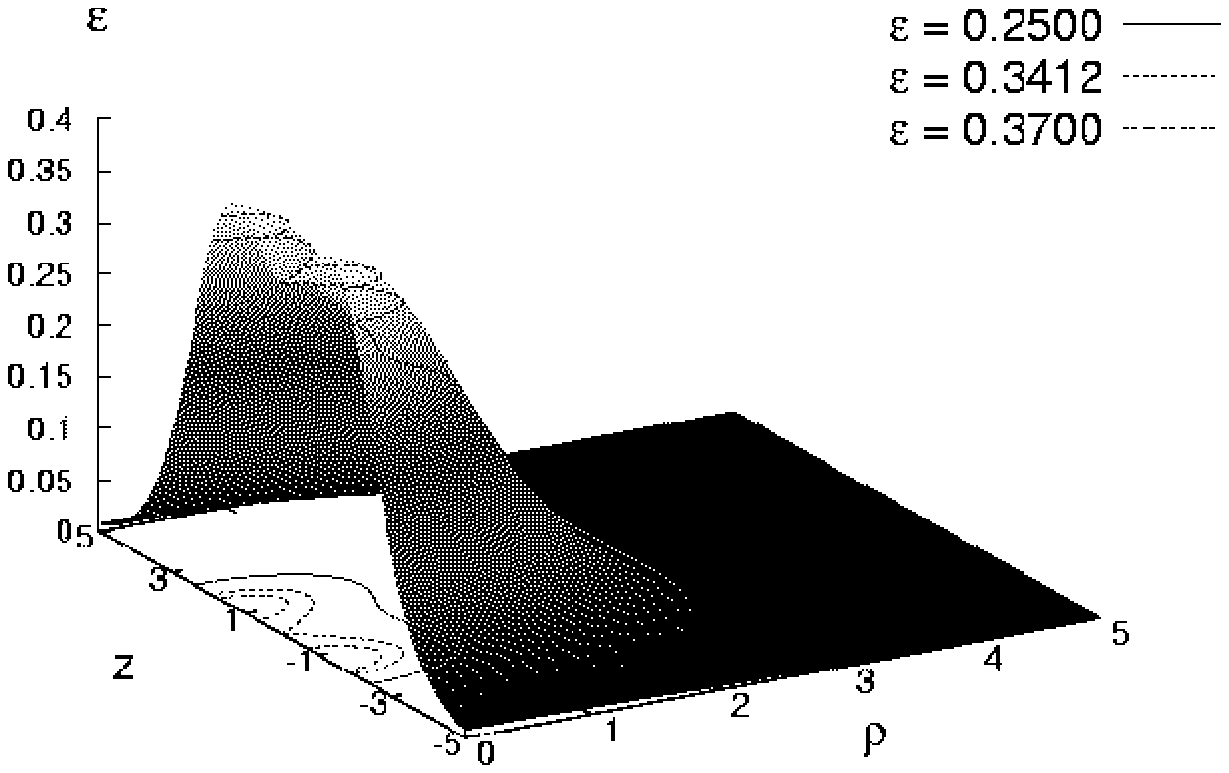}\\
(b)\epsfysize=4.5cm \epsffile{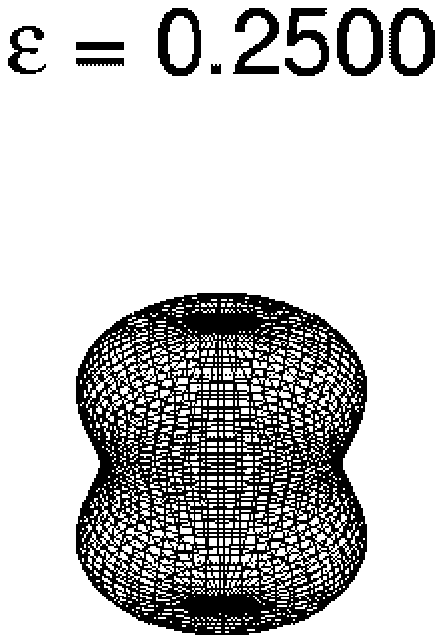}\hspace*{-0.cm}
\epsfysize=4.5cm \epsffile{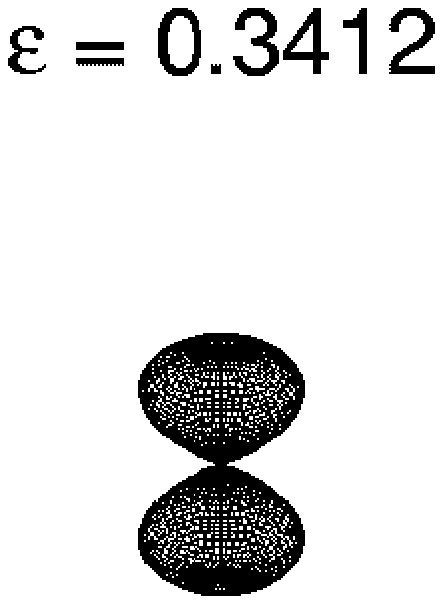}\hspace*{-0.cm}
\epsfysize=4.5cm \epsffile{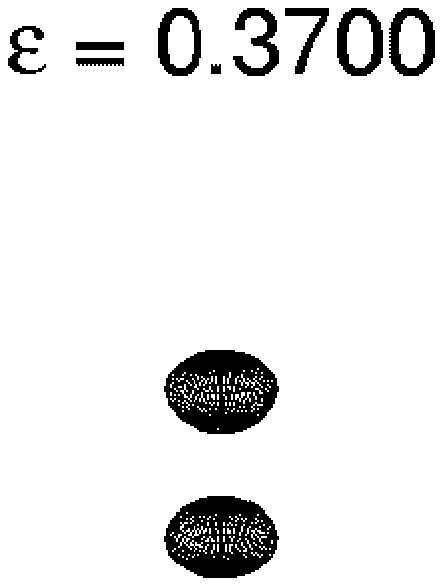}
\end{center}
{\bf Fig.~3} \small
(a) The energy density of the matter fields
for an electically charged monopole-antimonopole pair solution ($m=2$, $n=1$,
$\nu=0.32$, $\alpha=0.3$, $\beta=\gamma=0$).
(b) Also shown are surfaces of constant energy density.
\vspace{0.5cm}
}
\end{figure}

\section{\bf Rotating EYMHD black holes}

\subsection{\bf Non-Abelian Mass Formula}

We now derive the mass formula for 
stationary axially symmetric EYMHD black hole solutions
\begin{equation}
M = 2 \hat{T}\hat{S} + 2 \Omega J + \frac{D}{\gamma}
  +2\nu Q + 2{\tilde \psi}_{\rm el} Q (1-\varepsilon) 
, \label{namass2}
\end{equation}
where $\hat{T}$ and $\hat{S}$ are conveniently scaled dimensionless 
temperature and entropy, respectively.
The derivation is analogous to the derivation of the 
expressions for the mass and the angular momentum 
of the globally regular solutions.

Again we begin by recalling the general expressions \cite{wald}
for the global mass 
\begin{equation}
{\cal M} = {\cal M}_{\rm H}
 +\frac{1}{{4\pi G}} \int_{\Sigma}
 R_{\mu\nu}n^\mu\xi^\nu dV 
\ , \label{relmassy}
 \end{equation}
and the global angular momentum 
\begin{equation}
{\cal J} =  {\cal J}_{\rm H}
 -\frac{1}{{8\pi G}} \int_{\Sigma}
 R_{\mu\nu}n^\mu\eta^\nu dV 
\ , \label{relangmom} 
\end{equation}
where $\Sigma$ now denotes an asymptotically flat spacelike hypersurface 
bounded by the horizon ${\rm H}$,
and the horizon mass ${\cal M}_{\rm H}$ \cite{wald}
and the horizon angular momentum ${\cal J}_{\rm H}$ are given by 
\begin{eqnarray}
{\cal M}_{\rm H} & = & 
 - \frac{1}{{8\pi G}} \int_{\rm H} \frac{1}{2}
 \varepsilon_{\mu\nu\rho\sigma}\nabla^\rho \xi^\sigma dx^\mu dx^\nu
 = 2  {\cal T} {\cal S} 
 + 2 \frac{\omega_{\rm H}}{r_{\rm H}} {\cal J}_{\rm H}
\label{M} 
\\
 {\cal J}_{\rm H} & = &
 \frac{1}{16\pi G} \int_{\rm H} \frac{1}{2}
 \varepsilon_{\mu\nu\rho\sigma} \nabla^\rho \eta^\sigma dx^\mu dx^\nu
\ . 
\label{J} 
\end{eqnarray}
Substituting the horizon mass ${\cal M}_{\rm H}$ in Eq.~({\ref{relmassy})
and eliminating the horizon angular momentum ${\cal J}_{\rm H}$ 
yields for the global mass ${\cal M}$
\begin{equation}
{\cal M}= 2  {\cal T} {\cal S} 
        + 2 \frac{\omega_{\rm H}}{r_{\rm H}} {\cal J}
	+2 \left[
	   \frac{1}{8 \pi G} \int_{\Sigma} R_{\mu\nu}n^\mu\xi^\nu dV 
          +\frac{\omega_{\rm H}}{r_{\rm H}}\frac{1}{8 \pi G} 
                             \int_{\Sigma} R_{\mu\nu}n^\mu\eta^\nu dV 
	   \right]
  \ .
\end{equation}

Following now the same steps as for the globally regular solutions,
we obtain analogously to Eq.~(\ref{r-intIM}) and Eq.~(\ref{r-intIJ})
\begin{eqnarray}
{\DS {\cal M} - 2  {\cal T} {\cal S} - 2 \frac{\omega_{\rm H}}{r_{\rm H}} {\cal
J}- \frac{4\pi}{\kappa} {\cal D}
 = {\cal I}} &\equiv& {\DS - 4 \int_\Sigma  e^{2 \kappa \Psi} {\rm Tr}
 \left[  \left( F_{0\mu}+ \frac{\omega_{\rm H}}{r_{\rm H}} F_{\varphi\mu} \right
) F^{0\mu} \right]
 \sqrt{-g} dr d\theta d\varphi} \nonumber \\
 &&{\DS -  \int_\Sigma  {\rm Tr}\left[ \left( D_0 \Phi+ \frac{\omega_{\rm H}}{r_
{\rm H}} D_{\varphi}\Phi \right) D^0\Phi \right]
 \sqrt{-g} dr d\theta d\varphi} \ ,
\label{intb}
\end{eqnarray}
defining the integral ${\cal I}$.

We next note, that formally, we can express the integral ${\cal I}$
in terms of the integrals ${\cal I}_M$ Eq.~(\ref{r-intIM})
and ${\cal I}_J$ Eq.~(\ref{r-intIJ}), keeping in mind,
that $\Sigma$ is here bounded by the horizon.
Thus 
\begin{equation}
{\DS {\cal M} - 2  {\cal T} {\cal S} 
 - 2 \frac{\omega_{\rm H}}{r_{\rm H}} {\cal J}
 - \frac{4\pi}{\kappa} {\cal D}
 = {\cal I}_M-2 \frac{\omega_{\rm H}}{r_{\rm H}} {\cal I}_J }
\ . \label{tripleI}
\end{equation}

To evaluate ${\cal I}$,
we proceed again analogously to the globally regular case,
making use of the relations obtained for ${\cal I}_M$ and ${\cal I}_J$.
Analogously to Eq.~(\ref{intM}) and Eq.~(\ref{intJ}) we are then left with
\begin{equation}
{\cal I}=
 {\displaystyle 4 \int \left.
 {\rm Tr}  \left[  \left( A_0 + \frac{\omega_{\rm H}}{r_{\rm H}}
   \left( A_{\varphi} - W_\eta \right)  \right)
 e^{2 \kappa \Psi} F^{0r} \sqrt{-g} \right]
 \right|^\infty_{r_{\rm H}} d\theta d\varphi}
\ , \label{int} \end{equation}
where the integrand must be evaluated at the horizon
and at infinity.
Since the electrostatic potential 
is constant at the horizon (see Eqs.~(\ref{esp0})) and 
\begin{equation}
  \left. \left( A_0
 + \frac{\omega_{\rm H}}{r_{\rm H}} A_{\varphi} \right) \right|_{\rm H}
= \frac{\omega_{\rm H}}{r_{\rm H}} W_\eta
= - {\tilde \Psi}_{\rm el} \frac{\tau_z}{2}
\ , \label{esp2} \end{equation}
the integrand vanishes at the horizon,
and the only contribution to ${\cal I}$ comes from infinity.

At infinity both expressions have been evaluated in
Eq.~(\ref{r-IIfinal}). Thus the mass formula becomes
\begin{equation}
{\cal M} - 2  {\cal T} {\cal S}
 - 2 \frac{\omega_{\rm H}}{r_{\rm H}} {\cal J}
 - \frac{4\pi}{\kappa} {\cal D} = \frac{8\pi{\tilde \nu} Q}{e^2} 
 + \frac{8\pi {\tilde \Psi}_{\rm el}Q}{e} (1-\varepsilon)
\ . \label{Ifinal} \end{equation}
Returning again to dimensionless variables,
recalling Eq.~(\ref{psi}) and Eq.~(\ref{dimlessMJD}), and noting that
\begin{equation}
{\cal T} {\cal S} =
\frac{\sqrt{4\pi G}}{e\alpha G} TS \ , \ \ \
\frac{\omega_{\rm H}}{r_{\rm H}} =
\frac{e\alpha}{\sqrt{4\pi G}}  \Omega \ ,
\end{equation}
we obtain the mass formula Eq.~(\ref{namass2})
$$
\mu = 2 TS + 2 \Omega \zeta +\alpha^2 \frac{D}{\gamma} 
  +2\alpha^2\nu Q + 2\alpha^2{\tilde \psi}_{\rm el} Q (1-\varepsilon) \ ,
$$
or equivalently
$$
M = 2 \hat{T}\hat{S} + 2 \Omega J + \frac{D}{\gamma}
  +2\nu Q + 2{\tilde \psi}_{\rm el} Q (1-\varepsilon)  \ ,
$$
with scaled dimensionless temperature and entropy, 
$\hat{T}\hat{S} = TS/\alpha^2$.

This mass formula differs from the
EMD and EYMD mass formula Eq.~(\ref{namass}) in two respects.
First,
the last term is present only for magnetically neutral
black holes.
Second, the fourth term is an additional term,
not present for EMD and EYMD black holes. It appears for all
electrically charged EYMHD black holes,
and has the gauge potential parameter $\nu$ entering together
with the electric charge.
We note, that the first two terms and the last term 
do not appear in the mass formula Eq.~(\ref{Mreg})
for globally regular solutions.
Indeed, when the black hole horizon size is taken to zero, 
the first term vanishes, and the second and the last term cancel,
leaving the mass formula Eq.~(\ref{Mreg}) for globally regular solutions.

\subsection{\bf Numerical Results}

The numerical black hole calculations are performed analogously
to the calculations of globally regular solutions
\cite{schoen},
except that for black hole solutions we employ the
compactified dimensionless coordinate
$\bar x = 1-(x_{\rm H}/x)$, 
and we impose boundary conditions at the regular horizon.

For given coupling constants $\alpha$, $\beta$ and $\gamma$,
the rotating non-Abelian black hole solutions 
then depend on the horizon radius $x_{\rm H}$,
and on the rotational velocity of the horizon $\Omega$
in addition to the gauge potential parameter $\nu$
and the integers $m$ and $n$.

We exhibit in Fig.~4 an example of a dyonic rotating black hole,
which has $m=1$, $n=1$, horizon radius $x_{\rm H}=0.1$,
horizon angular velocity $\Omega=0.5$, gauge potential
parameter $\nu=0.04$
and the coupling constants $\alpha=0.3$, $\beta=0.1$, $\gamma=0.1$.
We again exhibit the energy density of the matter fields $\epsilon$,
Eq.~(\ref{edens}).

\begin{figure}[h!]
\noindent\parbox{\textwidth}{
\begin{center}
(a)\epsfysize=9.cm \epsffile{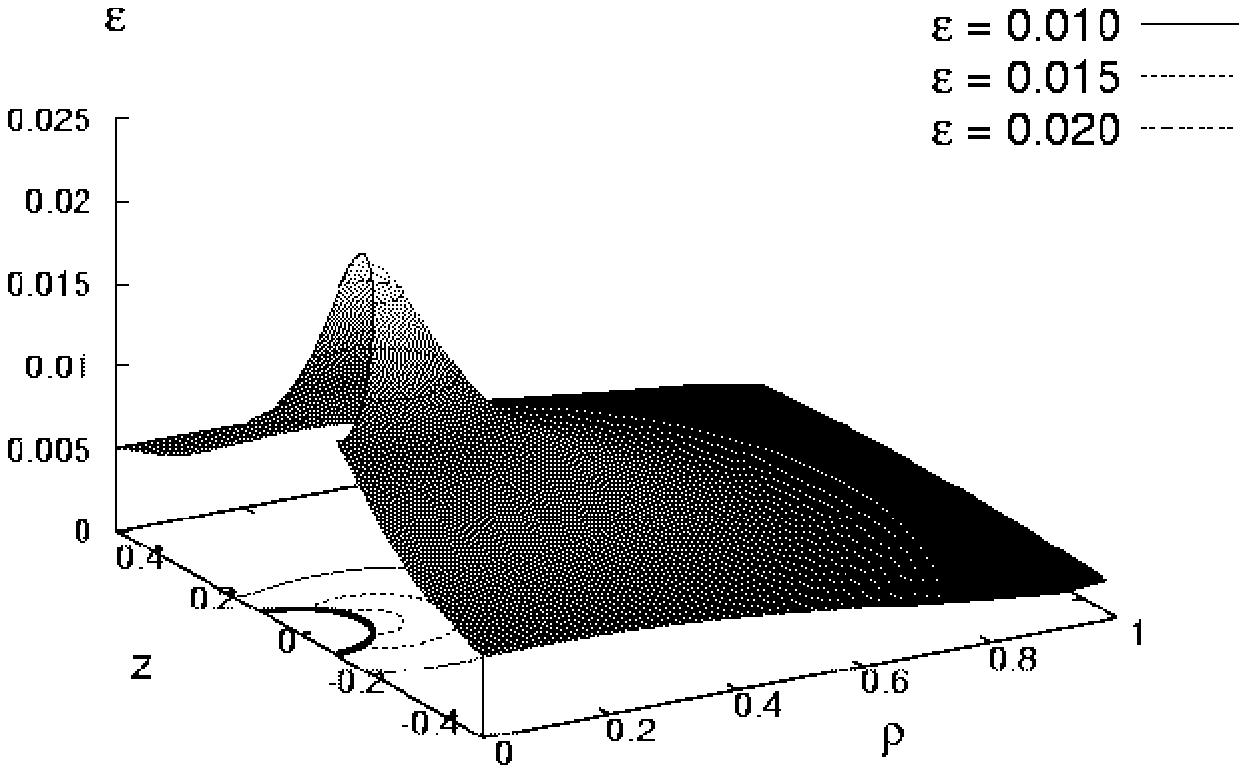}\\
(b)\epsfysize=4.5cm \epsffile{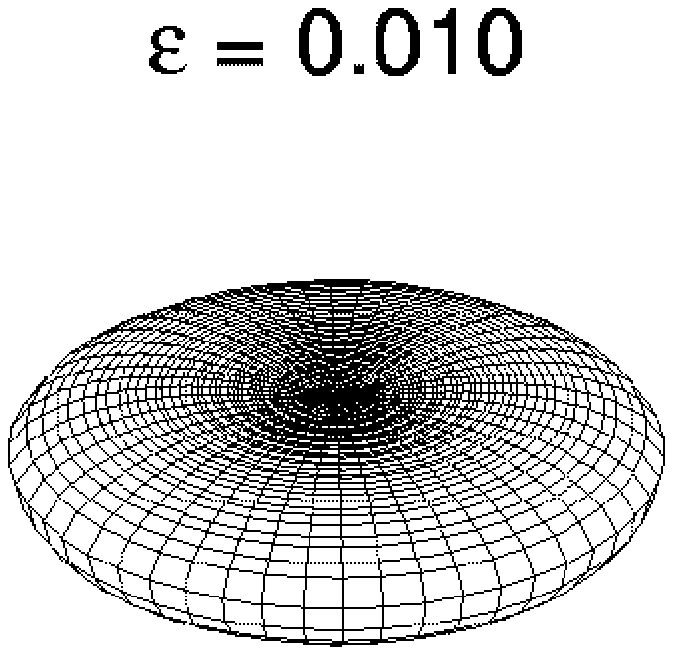}\hspace*{-0.cm}
\epsfysize=4.5cm \epsffile{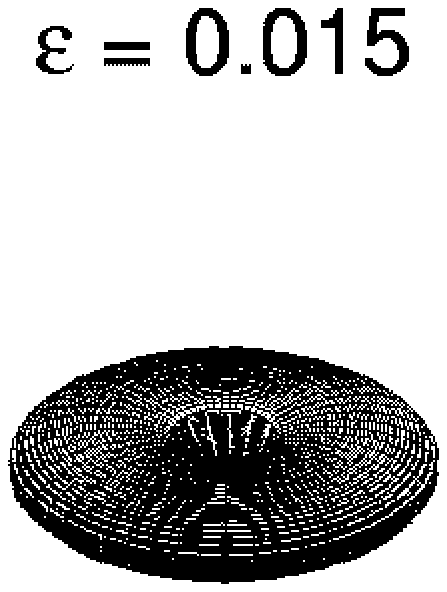}\hspace*{-0.cm}
\epsfysize=4.5cm \epsffile{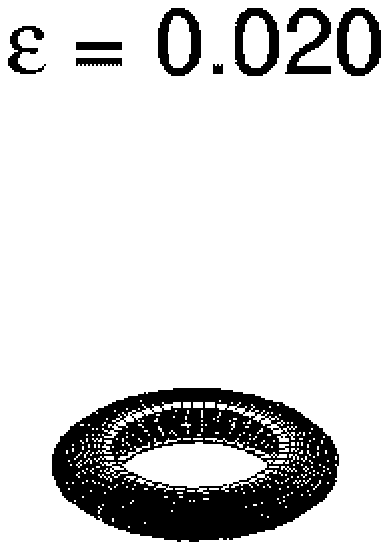}
\end{center}
{\bf Fig.~4} \small
(a) The energy density of the matter fields
for a dyonic rotating black hole ($m=1$, $n=1$, $x_{\rm H}=0.1$, $\Omega=0.5$,
$\nu=0.04$, $\alpha=0.3$, $\beta=0.1$, $\gamma=0.1$).
(b) Also shown are surfaces of constant energy density.
\vspace{0.5cm}
}
\end{figure}

\section{Conclusions}

We have considered non-perturbative globally regular
and black hole solutions of EYMHD theory.
These stationary axially symmetric solutions are asymptotically flat.
The solutions are characterized by two integers,
$m$ and $n$, related to the number of monopoles and antimonopoles
in the solutions,
and to the magnetic charge of the monopoles, respectively \cite{KKS}.
The black hole solutions carry non-Abelian hair outside
their regular horizon.

The globally regular solutions do not rotate, when they carry
a global magnetic charge. Only solutions with no global
magnetic charge can possess angular momentum,
which is then quantized in terms of the electric charge \cite{eugen}.
The globally regular solutions satisfy a simple mass formula
$$
M =  \frac{D}{\gamma} +2\nu Q \ .
$$
The presence of electric charge enforces stationarity
of the solutions, since it gives rise to an angular momentum density
(except for the spherical $n=1$ monopole).

The effective action of the globally regular solutions
can be expressed in terms
of the mass and the electric charge
$$
S^{\rm eff}= - \left(M - \nu Q\right)
= - \left(\frac{D}{\gamma} + \nu Q \right)\ .
$$
Based on the effective action we have shown,
that the mass of stationary solutions
can exhibit a ``loop'' close to the
maximal value of the coupling constant $\alpha_{\rm max}$, 
whereas the mass of static solutions can only exhibit a ``spike'' there
(for given $m$, $n$, $\beta$, $\gamma$ and $\nu$).

Rotating EYMHD black hole solutions 
satisfy the zeroth and the first law of black hole mechanics.
Here we have derived a non-Abelian mass formula 
for these black holes, which involves their
global charges and their horizon properties
$$
M = 2 \hat{T}\hat{S} + 2 \Omega J + \frac{D}{\gamma}
  +2\nu Q + 2{\tilde \psi}_{\rm el} Q (1-\varepsilon) \ .
$$
This mass formula differs from the
EMD and EYMD mass formula Eq.~(\ref{namass}),
since the last term is present only for magnetically neutral
black holes, 
and further an additional term is present for all 
electrically charged solutions,
where the gauge potential parameter $\nu$ is entering together
with the electric charge.
When the black hole horizon size is taken to zero, 
the mass formula for globally regular solutions is recovered.

Whether the presence of the dilaton also allows for a new uniqueness
conjecture for hairy black holes remains to be seen.
Clearly, when only the mass, the angular momentum
and the electric and magnetic charges are considered,
the black hole solutions are not uniquely determined by these 
global charges.

In the numerical calculations we have only began to
investigate the large parameter space for the black hole solutions.
Here further investigations might reveal new phenomena,
not encountered previously for non-Abelian black holes.
For instance, non-Abelian counterexamples to
the staticity theorem might arise
as well as counterrotating black holes \cite{KKN-c}.
Also, systems of non-Abelian black holes with regular non-degenerate
horizons might exist.

\begin{acknowledgments}
BK gratefully acknowledges support by the German Aerospace Center,
FNL by the Ministerio de Educaci\'on y Ciencia 
under grant EX2005-0078.
\end{acknowledgments}


\end{document}